\documentclass[floats,floatfix,showpacs,amssymb,prd,twocolumn,superscriptaddress,nofootinbib]{revtex4-1}
\usepackage{amssymb,amsmath,verbatim,mathtools,needspace,enumitem,etoolbox,graphicx,physics,microtype}

\usepackage[dvipsnames, usenames]{xcolor}

\definecolor{linkcolor}{rgb}{0.0,0.3,0.5}
\definecolor{urlcolor}{rgb}{0.27,0.55,0.}
\definecolor{funcolor}{rgb}{0.65, 0.16, 0.16}
\usepackage[unicode, colorlinks=true, linkcolor=linkcolor, citecolor=linkcolor, filecolor=linkcolor,urlcolor=linkcolor, pdfusetitle]{hyperref}
\usepackage[all]{hypcap}
\usepackage{stackengine,xcolor,scalerel}

\usepackage[T1]{fontenc}
\usepackage[utf8]{inputenc}
\usepackage[normalem]{ulem}

\allowdisplaybreaks
\def\apj{\rm{ApJ}}
\def\apjl{\rm{ApJ}}

\def\aap{\rm{A\&A}}
                                 
\def\mnras{\rm{MNRAS}}
\def\prd{\rm{PRD}}
\def\prl{\rm{PRL}}

\def\nat{\rm{Nature}}
\def\cqg{\rm{CQG}}
\def\grg{\rm{GRG}}
\def\lrr{\rm{LRR}}

\newcommand{\yanbei}[1]{{\textcolor{black}{#1}}}
\newcommand{\red}[1]{{\textcolor{black}{#1}}}

\newcommand{\caltech}{\affiliation{TAPIR 350-17, California Institute of Technology, 1200 E California Boulevard, Pasadena, CA 91125, USA}}
\newcommand{\bham}{\affiliation{School of Physics and Astronomy and Institute for Gravitational Wave Astronomy, University of Birmingham, Birmingham, B15 2TT, UK}}

\begin{document}

\title{Optimizing LIGO with LISA forewarnings to 
improve black-hole spectroscopy}

\author{Rhondale Tso} \email{rtso@caltech.edu} \caltech
\author{Davide Gerosa} 
\thanks{Einstein Fellow}
\email{dgerosa@caltech.edu} \caltech \bham
\author{Yanbei Chen} \email{yanbei@caltech.edu} \caltech

\date{\today}

\begin{abstract}

The early inspiral of massive stellar-mass black-hole binaries merging in LIGO's sensitivity band will be detectable at low frequencies by the upcoming space mission LISA. LISA will predict, with years of forewarning, the time and frequency with which binaries will be observed by LIGO.  We will, therefore, find ourselves in the position of knowing that a binary is about to merge, with the unprecedented opportunity to optimize ground-based operations to increase their scientific payoff. 
We apply this idea to detections 
of multiple ringdown modes, or black-hole spectroscopy. %
Narrowband tunings can  
boost the detectors' sensitivity at frequencies corresponding to the first subdominant ringdown mode and largely improve our prospects to 
experimentally test
the Kerr nature of astrophysical black holes. We define a new consistency parameter between the different modes, called $\delta {\rm GR}$, and show that, in terms of this measure, optimized configurations have the potential to double the effectiveness of black-hole spectroscopy when compared to standard broadband setups.

\end{abstract}

\maketitle

\section{Introduction}
The first detection of merging black-hole (BH) binaries by the LIGO ground-based detectors is one of the greatest achievements in modern science. \red{Some of the binary  component} masses are as large as $\sim\!30 M_\odot$ and unexpectedly exceed those of all previously known stellar-mass BHs \cite{2018arXiv181112907T}. These  systems might also be visible by the future spaced-based detector LISA, %
which will soon observe the gravitational-wave (GW) sky in the mHz regime~\cite{2017arXiv170200786A}.  LISA will measure the early inspiral stages of BH binaries predicting, with years to weeks of forewarning, the time at which the binary will enter the LIGO band \cite{2016PhRvL.116w1102S}. This will allow electromagnetic observers to concentrate on the source's sky location, thus %
increasing the likelihood of observing counterparts.  Multiband GW observations have the potential to shed light on BH formation channels \cite{2016PhRvD..94f4020N,2016ApJ...830L..18B,2017MNRAS.465.4375N,2017ApJ...842L...2C,2017PhRvD..96f3014I,2018arXiv180208654S, 2019PhRvD..99j3004G}, constrain dipole emission \cite{2016PhRvL.116x1104B}, enhance searches and 
parameter estimation~\cite{2016PhRvL.117e1102V,2018PhRvL.121y1102W}, and provide new measurements of the cosmological parameters \cite{2017PhRvD..95h3525K,2018MNRAS.475.3485D}.

Here we explore the possibility of improving the science return of ground-based GW observations by combining LISA forewarnings to active interferometric techniques. LISA observations of stellar-mass BH binaries at low frequencies can be exploited to prepare detectors on the ground in their most favorable configurations for a targeted measurement. Optimizations can range from the most obvious ones (for instance just \yanbei{ensuring} the detectors are operational), to others that require more experimental work, like changing the input optical power, modifying mirror transmissivities and cavity tuning phases, and changing the squeeze factor and angle of the injected squeeze vacuum (see, e.g.,~\cite{2014RvMP...86..121A}).  Tuning the optical setup of the interferometer can allow to boost the signal-to-noise ratio (SNR) of specific features of the signal ``on demand'' (only at the needed time, only at the needed frequency).

In particular, we apply this line of reasoning to the so-called \emph{black-hole spectroscopy}: testing the nature of BHs through their ringdown modes. Narrowband tunings were previously explored \yanbei{for studying} the detectability of neutron-star mergers  \cite{2002PhRvD..66j2001H,2018PhRvD..98d4044M,2019PhRvD..99j2004M}
 and stochastic backgrounds \cite{2018CQGra..35l5002T}, and are here \yanbei{proposed for} BH science for the first time.

The perturbed BH resulting from a merger vibrates at very specific frequencies. These {quasi-normal modes} of oscillation are damped by GW emission, resulting in the so-called BH ringdown \cite{1970Natur.227..936V,2009CQGra..26p3001B}. If BHs are described by the Kerr solution of General Relativity (GR) \cite{1963PhRvL..11..237K}, all these resonant modes are allowed to depend on two quantities only: mass and spin of the perturbed BH \cite{1968CMaPh...8..245I,1971PhRvL..26..331C,1996bhut.book.....H}. This is a consequence of the famous \emph{no-hair theorems}: as two BHs merge, all additional complexities (hair) of the spacetime 
are dissipated away in GWs, and a Kerr BH is left behind.
The detection of frequency and decay time of one  quasi-normal mode can therefore be used to infer mass and spin of the post-merger BH. Measurements of each additional mode provide consistency tests of the theory. This is the main idea behind BH spectroscopy: much like atoms' spectral lines can be used to identify nuclear elements and test quantum mechanics, quasi-normal modes can be used to probe the nature of BHs and test GR \cite{1980ApJ...239..292D,2004CQGra..21..787D,2006PhRvD..73f4030B,2018GReGr..50...49B}.
Despite its elegance, BH spectroscopy turns out to be challenging in practice as it requires loud GW sources and improved data analysis techniques \cite{2016PhRvL.117j1102B,2017PhRvD..95j4026M,2018PhRvD..97d4048B,2017PhRvL.118p1101Y,2018PhRvD..97j4065B,2018PhRvD..98h4038B}. 

\begin{figure}
\includegraphics[width=\columnwidth]{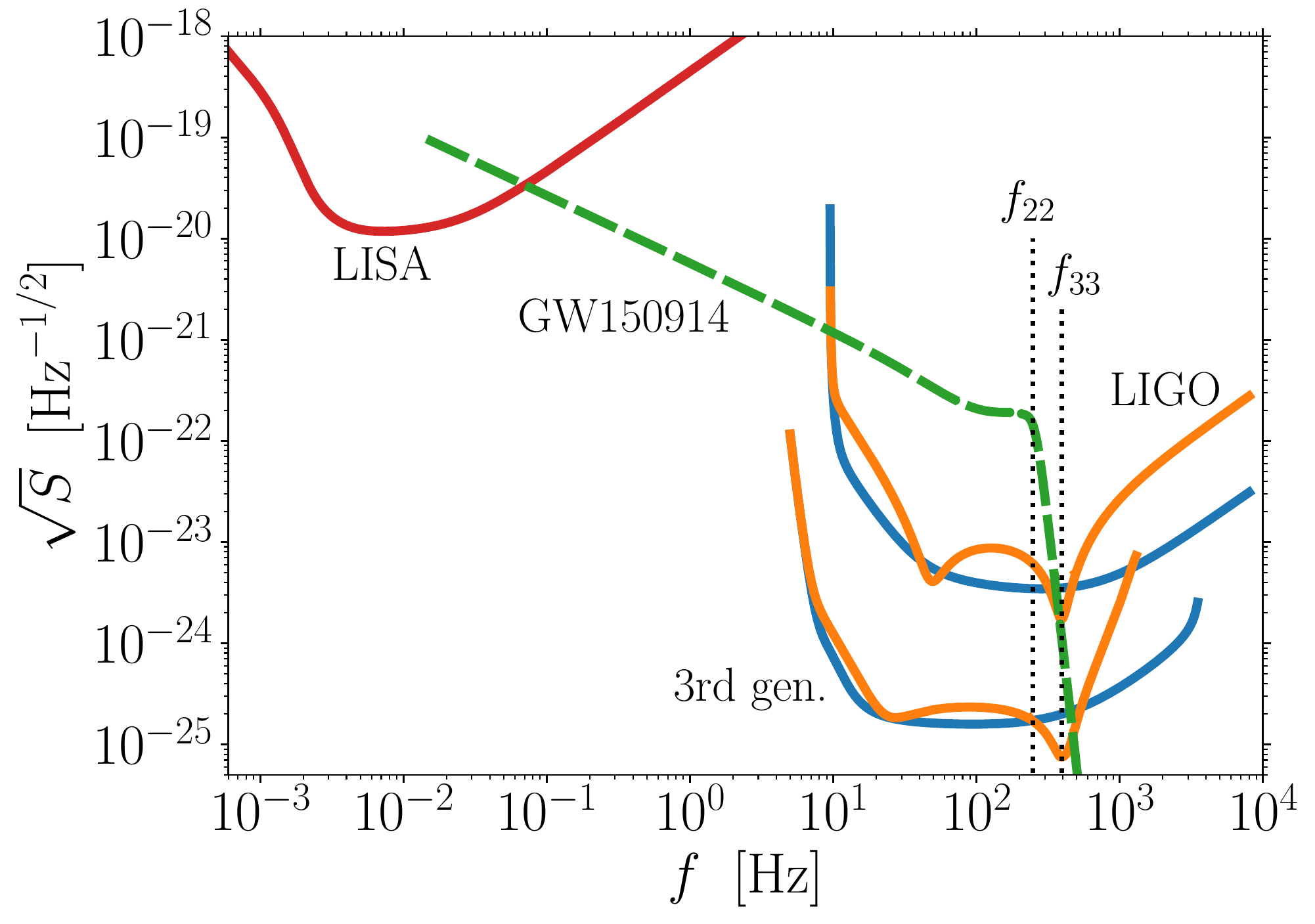}
\caption{GW amplitude $\sqrt{S_h}=2| \tilde{h} | \sqrt{f}$ of a black-hole binary source similar to GW150914 compared to the noise curves $\sqrt{S_n}$ of LISA  \cite{2018arXiv180301944C}, LIGO \red{\cite{2016LRR....19....1A}, and a planned 3rd-generation detector \cite{2017CQGra..34d4001A} (both in their broadband configurations and with narrowband tunings). } Optimized {narrowbanding}
enhances (decreases) the detector sensitivity around the frequency $f_{33}$ ($f_{22}$) of the first subdominant (dominant) %
 mode of the BH ringdown.
The BH binary waveform is generated using the approximant of \cite{2016PhRvD..93d4007K} with $m_1+m_2=65M_\odot$, $q=0.8$, $D=410$ Mpc, $\iota=150^\circ$ assuming optimal orientation ($\theta=\phi=\psi=0$).}
\label{idea}
\end{figure}

The main idea behind our study is illustrated in Fig.~\ref{idea}. A GW source like GW150914 emits GWs at $\sim\!0.1$ Hz  and is visible by LISA with SNR$\sim\!5$. After $\sim\!10$ years, the emission frequency  reaches $\sim\!10$ Hz and the source appears in the \red{sensitivity band of LIGO or a future ground-based detector}. The excitation amplitude of the dominant 
quasi-normal mode is $\sim\!10$ times
 higher than the first subdominant mode. 
 The latter is likely going to be too weak to perform BH spectroscopy.
{Optimized narrowband tunings} can boost the detectability of the weaker mode at the expense of the rest of the signal, making BH spectroscopy possible.

This paper is organized as follows. In Secs.~\ref{bhspecsec} and \ref{tuning} we introduce BH spectroscopy and narrowband tunings, respectively. Our results are illustrated in Sec.~\ref{secresults}. We draw our conclusions in Sec.~\ref{secconcl}. Hereafter, we use geometric units $c=G=1$.

\section{Black-hole spectroscopy}
\label{bhspecsec}

\subsection{Black-hole ringdown}
Let us consider a perturbed BH with detector-frame mass $M$ and dimensionless spin $j$.
GW emission during ringdown can be described by a superposition of damped sinusoids, labeled by $l\geq 2$, $0\leq |m| \leq l$ and $n\geq 0$ \cite{1973ApJ...185..635T}. For simplicity, we only  consider the fundamental overtone $n=0$.  

Each mode is described by its frequency $\omega_{lm}$ and decay time $\tau_{lm}$. The GW strain can be written as \cite{2007PhRvD..76j4044B,2012PhRvD..85b4018K},
\begin{eqnarray}
 h (t) &=& \sum_{l,m>0} B_{lm} e^{-t/\tau_{lm}} \cos \left( \omega_{lm} t + \gamma_{lm} \right)\,,
\\
B_{lm} &=& \frac{\alpha_{lm} M }{ D} \sqrt{ \left( F_+ Y^{lm}_+ \right)^2 + \left( F_\times Y^{lm}_\times \right)^2  }\,, \\
 \gamma_{lm} & =& \phi_{lm} + m \beta + \arctan \left(  \frac{F_\times Y^{lm}_\times }{ F_+ Y^{lm}_+ }  \right)   \,,
\\
Y_{+, \times}^{lm} (\iota) &=& _{-2}Y_{lm}(\iota,\beta\!=\!0) \pm (-1)^l _{-2}Y_{l-m}(\iota,\beta\!=\!0)\,,
\end{eqnarray} 
where $\mathcal{\alpha}_{lm}$ and $\phi_{lm}$ are the mode amplitudes and phases, $D$ is the luminosity distance to the source, $_{-2}Y_{lm}(\iota,\beta)$ are the spin-weighted spherical harmonics, $F_{+,\times} (\theta, \phi, \psi)$ are the single-detector antenna patterns \cite{1987thyg.book..330T}. The angles $\iota$ and $\beta$ describe the orientation of the BH, with $\iota$ ($\beta$) being the polar (azimuthal) angle \yanbei{of the wave propagation direction} measured with respect to the BH spin axis. 
In the conventions of \cite{1989PhRvD..40.3194E,1992PhRvD..46.5236F}, the frequency-domain strain reads,
\begin{equation}\label{FTresult}
\tilde{h}(f) = \sum_{l,m>0} B_{lm} \frac{ -\omega_{lm} \sin \gamma_{lm} + ( 1 / \tau_{lm} - i \omega ) \cos  \gamma_{lm}   }{ \omega_{lm}^2 - \omega^2 + 1 / \tau_{lm}^2 - 2 i \omega  / \tau_{lm} }\,,
\end{equation}
where $f= \omega /2 \pi$ is the GW frequency. 

The dominant mode corresponds to $l$=2, $m$=2 (hereafter ``22''), while the first subdominant is usually $l$=3, $m$=3 (hereafter ``33'').  Other modes %
might sometimes be stronger than the 33 mode for specific sources. For instance, the 33-mode is suppressed for $q \simeq 1$ or  $\sin\iota\simeq 0$ (e.g \cite{2014PhRvD..90l4032L,2016PhRvD..94h4024B,2018PhRvD..97d4048B}). 
Here we perform a simple two-mode analysis considering the 22 and 33 modes only. Strictly speaking, the ringdown modes have angular distributions described by spheriodal, instead of spherical harmonics.  However, for the final black-hole spins we consider, the 22 and 33 spin-weighted spherical harmonics have more than 99\% overlap with the corresponding spin-weighted spheroidal harmonics~\cite{2014PhRvD..90f4012B,2019arXiv190308284G}, which  is accurate enough for this study.\footnote{We do note that, for the final black-hole spins we are considering, $_{-2}S_{22}$ and $_{-2}Y_{32}$  have overlap between 0.05 and 0.1, which does cause the 22 ringdown mode to show up significantly in the spherical-harmonic mode $h_{32}$.  This is nevertheless consistent with the 99\% overlap between  $_{-2}Y_{22}$ and $_{-2} S_{22}$, because $\sum_{l'}|\langle _{-2} Y_{l'm} | _{-2} S_{lm}\rangle |^2=1  $. }
For simplicity, we restrict ourselves to non-spinning binary BHs with source-frame masses $m_1$ and $m_2$; we address the impact of this assumption in Sec.~\ref{secconcl}. Redshifted masses $m_i(1+z)$ are computed from the luminosity distance $D$ using the Planck cosmology \cite{2016A&A...594A..13P}. Mass $M$ and spin $j$ of the post-merger BH are estimated using fits to numerical relativity simulations \cite{2012ApJ...758...63B,2009ApJ...704L..40B} as implemented in Refs.~\cite{2016PhRvD..93l4066G}.  Quasi-normal frequencies $\omega_{lm}$ and decay times $\tau_{lm}$ are estimated from \cite{2006PhRvD..73f4030B}. %
We estimate the excitation amplitudes $\alpha_{lm}$ given the mass ratio $q=m_2/m_1\leq 1$ of the merging binary using the expressions reported by \cite{2012PhRvD..85b4018K}. BH ringdown parameter estimation has been shown to depend very weakly on the phase offsets $\phi_{lm}$ \cite{2006PhRvD..73f4030B}, which we thus set to 0 for simplicity (cf. also \cite{2008PhRvD..78d4046B}).

 \subsection{Waveform model and GR test}
 In BH spectroscopy, one assumes that quasi-normal modes frequencies $\omega_{lm}$ and decay times $\tau_{lm}$ \yanbei{for different modes} depend separately on $M$ and $j$, and then look for consistencies between the different estimates.\footnote{For simplicity we only vary $\omega_{lm}$ and $\tau_{lm}$ while keeping $\alpha_{lm}$ fixed to their GR values.} Considering the $22$ and $33$ modes only, one can write the waveform as,
 \begin{equation} 
  h = h_{22} ( M_{22}, j_{22} ) + h_{33} ( M_{33}, j_{33} )
  \end{equation}
  and use data to estimate the parameters,
  \begin{equation} 
 \boldsymbol{\lambda} \equiv  \{ M_{22}, j_{22}, M_{33}, j_{33}\}\,.
  \end{equation}
Deviations from GR may cause non-zero values of,
  \begin{equation}
  \label{defepsilon}
 \epsilon_M  \equiv \frac{M_{22}-M_{33}}{(M_{22}+M_{33})/2}\,,\quad  \epsilon_ j  \equiv \frac{j_{22}-j_{33}}{(j_{22}+j_{33})/2}  \,.
  \end{equation}
We, therefore, seek to maximize our ability to estimate $\epsilon_M$ and $\epsilon_j$ from the observed data. 

{Given true values $\bar\lambda_i$, each independent noise realization will result in estimates $\tilde \lambda_i$ given by,
\begin{equation}
\tilde \lambda_i = \bar\lambda_i + \delta \lambda_i \, , 
\end{equation}
where $\delta \lambda_i$ are random variables driven by noise fluctuations in a way that depends on both the signal and the estimation scheme.  Measured values of deviation from GR can be obtained by inserting measured values $\tilde M_{22,33}$ and $\tilde j_{22,33}$ into Eq.~\eqref{defepsilon}, resulting in,
 \begin{equation}
 \tilde\epsilon_M  = \frac{\tilde M_{22}-\tilde M_{33}}{(\tilde M_{22}+\tilde M_{33})/2}\,,\quad  \tilde \epsilon_ j  = \frac{\tilde j_{22}-\tilde j_{33}}{(\tilde j_{22}+\tilde j_{33})/2}  \,.
  \end{equation}
At linear order one gets $\tilde \epsilon_M = \bar\epsilon_M +\delta\epsilon_M$ and $\tilde \epsilon_j = \bar\epsilon_j +\delta\epsilon_j$, with,
\begin{align}
\label{deltaepsilon}
\delta\epsilon_M\!=\!\frac{\bar M_{33} \delta M_{22} -\bar M_{22} \delta M_{33}}{(\bar M_{22}+\bar M_{33})^2/4}\,, \quad \delta\epsilon_j \!=\! \frac{\bar j_{33} \delta j_{22} -\bar j_{22} \delta j_{33}}{(\bar j_{22}+\bar j_{33})^2/4}\,.\quad 
\end{align}
In the absence of any deviations from GR, one has $\bar M_{22}=\bar M_{33} =\bar M$ and $\bar j_{22} =\bar j_{33} =\bar j$,  but $\epsilon_M$ and $\epsilon_j$ will \yanbei{have statistical fluctuations} given by,
\begin{equation}
\label{deltaepsilonapp}
\delta \epsilon_M =\frac{\delta M_{22}-\delta M_{33}}{\bar M}\,,\quad 
\delta \epsilon_j =\frac{\delta j_{22}-\delta j_{33}}{\bar j}\,.
\end{equation}
The levels of these fluctuations will quantify our ability to test GR.   \yanbei{In fact, Eqs.~\eqref{deltaepsilonapp} are good approximations to \eqref{deltaepsilon}, as long as fractional deviation from GR is small,  i.e., when $\bar\epsilon_M\ll1$, and $\bar\epsilon_j \ll 1$. 
 }

\subsection{Estimation errors}

{ The covariance matrix $\sigma_{ij}$, namely the expectation values,
\begin{equation}
\sigma_{ij} \equiv \langle \delta\lambda_i\delta\lambda_j\rangle
\end{equation}
can be bounded by the Fisher information formalism~\cite{1994PhRvD..49.2658C} (but see  \cite{2013PhRvD..88h4013R}). The conservative bound for the error is given by the inverse of the Fisher Information matrix:
\begin{equation}
\sigma_{ij} = \mathbf\Gamma_{ij}^{-1} \,,\quad 
\mathbf\Gamma_{ij} = \bigg(
\frac{\partial \tilde h }{ \partial \lambda_i} \bigg| \frac{ \partial \tilde h}{  \partial \lambda_j}\bigg), 
\end{equation}
where parentheses indicate the standard noise-weighted inner product.
}

{In our case, the covariance matrix can be broken  into blocks, }
\begin{equation}\label{fisher}
\mathbf{\Gamma}^{-1} = 
\begin{bmatrix}
(\mathbf{\Gamma}^{-1})_{2222} & (\mathbf{\Gamma}^{-1})_{2233} \\ %
(\mathbf{\Gamma}^{-1})_{3322} & (\mathbf{\Gamma}^{-1})_{3333}
\end{bmatrix}
\end{equation}
corresponding to the couples $(M_{22}, j_{22})$ and  $(M_{33}, j_{33})$. {The diagonal block $(\mathbf\Gamma^{-1})_{2222}$  corresponds to errors when estimating  $(M_{22}, j_{22})$  alone \yanbei{(marginalizing over other uncertainties)}, the diagonal block $(\mathbf\Gamma^{-1})_{3333}$  corresponds to errors when estimating  $(M_{33}, j_{33})$  alone  \yanbei{(marginalizing over other uncertainties)},  while the non-diagonal blocks contain error correlations.  }

From the covariance matrix \yanbei{for $(M_{22},j_{22},M_{33},j_{33})$}, one obtains the following expectation values, %
\begin{align}
\langle\delta\epsilon_M ^2\rangle & = \frac{\displaystyle  \sigma_{M_{22}M_{22}}\! -\!2 \sigma_{M_{22} M_{33}} +\sigma_{M_{33}M_{33}}}{\bar M^2} \, , \\
\langle\delta\epsilon_j ^2\rangle & = \frac{\displaystyle  \sigma_{j_{22}j_{22}}\! -\!2 \sigma_{j_{22} j_{33}} +\sigma_{j_{33}j_{33}}}{\bar j^2} \, , \\
\langle\delta\epsilon_M \delta\epsilon_j\rangle & =
\frac{\sigma_{M_{22} j_{22}} -\sigma_{M_{33} j_{22} }-\sigma_{j_{22}M_{33}} +\sigma_{M_{33} j_{33}} }{\bar M \bar j }\,.
\end{align}
\yanbei{which are elements of the covariance matrix of $(\delta\epsilon_M,\delta\epsilon_j)$. } For concreteness, we define a scalar figure of merit,
\begin{equation}\label{dGR}
\delta{\rm GR} = 
\begin{vmatrix} 
\langle \delta \epsilon_M^2  \rangle & \langle \delta \epsilon_M \delta \epsilon_j \rangle
\\
\langle \delta \epsilon_M \delta \epsilon_j \rangle & \langle\delta\epsilon_j^2\rangle  
\end{vmatrix} ^{1/4}
\end{equation}
to quantify our ability to test GR. 
More specifically, $\delta{\rm GR}$ measures our statistical error in revealing deviations from GR. One has the strongest possible test of GR when  $\delta{\rm GR} \rightarrow 0$, corresponding to $\mathbf{\Gamma}^{-1} \rightarrow \mathbf{0}$, in which case any deviation from GR will be revealed with vanishing statistical error.  Large values of $\delta{\rm GR}$ would require larger deviations from GR \yanbei{[i.e., larger true values of $(\epsilon_M,\epsilon_j)$]} in order to be detectable.%

Given values of $\delta{\rm GR}$ from both a design and an optimized detector configuration, it is useful to define the narrowband gain,
\begin{equation}
\zeta = \frac{\delta{\rm GR}^{(\rm Design)} - \delta{\rm GR}^{(\rm Optimized)}}{\delta{\rm GR}^{(\rm Design)}}\,,
\label{gain}
\end{equation}
where $\zeta\!=\!1$ ($\zeta\!=\!0$) means that the narrowbanding procedure is maximally effective (irrelevant).

\subsection{Error correlations between modes}%

We note that the 22-33 correlation components of the Fisher information matrix, as well as its inverse, are expected to be small because the %
two 
modes are well separated in the frequency domain. In particular, $\partial h(\omega)/\partial M_{22}$ and $\partial h(\omega)/\partial j_{22}$ peak near $\omega_{22}$ with widths $\sim\!1/\tau_{22}$, while $\partial h(\omega)/\partial M_{33}$ and $\partial h(\omega)/\partial j_{33}$ peak near $\omega_{33}$ with widths $\sim\!1/\tau_{33}$.
For this reason, the pairs $(\delta M_{22}, \delta j_{22})$  and  $(\delta M_{33}, \delta j_{33})$ are nearly statistically independent from each other. Estimation error for $\epsilon_M$ and $\epsilon_j$ can be viewed as \yanbei{(almost)} independently contributed from the 22 and 33 modes and summed by quadrature.   One has, approximately, 
\begin{align}
\langle\delta\epsilon_M ^2\rangle & \approx \frac{\displaystyle  \sigma_{M_{22}M_{22}} +\sigma_{M_{33}M_{33}}}{\bar M^2} \, ,\\
\langle\delta\epsilon_j ^2\rangle & \approx  \frac{\displaystyle  \sigma_{j_{22}j_{22}} +\sigma_{j_{33}j_{33}}}{\bar j^2} \, , \\
\langle\delta\epsilon_M \delta\epsilon_j\rangle & \approx 
\frac{\sigma_{M_{22} j_{22}}  +\sigma_{M_{33} j_{33}} }{\bar M \bar j }\,.
\end{align}
In other words, the covariance matrix of $(\delta\epsilon_M,\delta \epsilon_j)$ is approximated by the sum of those of $(\delta M_{22}/ \bar M,\delta j_{22}/\bar j)$ and $(\delta M_{33}/ \bar M,\delta j_{33}/\bar j)$. 

We quantify this claim by calculating values $\delta{\rm GR}$ where the off-diagonal sub-matrices $(\mathbf{\Gamma}^{-1})_{3322}$ and $(\mathbf{\Gamma}^{-1})_{2233}$ are artificially set to zero. For the population of sources studied in Sec.~\ref{popsec}, and observed by LIGO, the median difference between the two estimates is as small as $1.6\%$ ($4.0\%$) for broadband (narrowband) configurations.

For this reason, some insight can be gained by visualizing the error region in the $(M_{22}, j_{22})$ and $(M_{33}, j_{33})$ planes separately (c.f Sec.~\ref{singlesources})\yanbei{: errors} in $(\delta \epsilon_M ,\delta \epsilon_j)$ are well approximated by the quadrature sum of errors indicated by those regions. We stress however, that correlations are fully included in all values of $\delta {\rm GR}$ reported in the rest of this paper.

\section{Narrowband tunings} \label{tuning}
As an example of a possible narrowband setup,  we consider the detuning of the signal-recycling cavity (cf.~\cite{2018PhRvD..98d4044M,2018CQGra..35l5002T} where a similar setup was also explored).
Second-generation GW detectors make use of signal-recycling optical configurations (or {resonant side-band extraction}) \cite{1988PhRvD..38.2317M,1996PhLA..217..305H,2014ASSL..404...57V}. A signal-recycling mirror is placed at the dark port of a Fabry-Perot Michelson interferometer, which is the configuration used in first-generation detectors. The transmittance  $T_{\rm SRM}$ of this mirror determines the fraction of signal light which is sent back into the arms, possibly with a detuning phase $\phi_{\rm SRM}$. Both these parameters affect the optical resonance properties of the interferometer~\cite{1988PhRvD..38.2317M,1996PhLA..217..305H}, as well as its optomechanical dynamics~\cite{2002PhRvD..65d2001B,2003PhRvD..67f2002B}.  Together with the homodyne readout phase $\phi_\mathrm{hd}$,  $T_{\rm SRM}$ and   $\phi_{\rm SRM}$ are responsible for the quantum noise spectrum of the interferometer, allowing for noise suppression near  optical and optomechanical resonances %
\cite{2001PhRvD..64d2006B}. 

In this paper, we consider narrowbanding of both LIGO in its design configuration and future 3rd-generation detectors.
The LIGO design
noise curve is a finalized experimental setup which allows us to perform a
focussed assessment of the impact of narrowbanding onto BH spectroscopy over a large number of sources. However, more sensitive ground-based interferometers are currently being planned and are expected to be operational by the 2030s \cite{2010CQGra..27s4002P,2017CQGra..34d4001A}. 
Multiband observations and LISA forewarnings might happen with a network of ground-based detectors perhaps 10 times more sensitive than LIGO.

In order to select the best detuned configuration to perform BH spectroscopy, one needs to choose values of $(T_{\rm SRM},\phi_{\rm SRM},\phi_{\rm hd})$ that %
boost sensitivity around the 33 frequency. %
For LIGO, we generate $60^3$ noise curves with equal spacing  in $\phi_\mathrm{SRM} \in [-0.12\pi,0.12\pi]$, $T_{\rm \tiny SRM} \in [0.001,0.2]$, and $\phi_\mathrm{hd} \in [0, \pi]$. This parameter space is capable of capturing the central frequencies of both the 22 and 33 mode for binaries with $q\in[0.2\!-\!0.9]$ and total masses $m_1+m_2\in[20 M_\odot \!-\!100 M_\odot]$.  
Noise  curves are generated using  pyGWINC  \cite{gwinc}.
The LIGO design configuration corresponds to $T_{\rm SRM}=0.2$, $\phi_\mathrm{SRM}=0$, and $\phi_\mathrm{hd}=\pi/2$. The broadband noise curves reported by \cite{2016LRR....19....1A,LIGOcurve} are reproduced within $\Delta \log\!{S_n}/\log\!S_n \lesssim 0.2\%$  throughout the entire frequency band. %
For each given source, we select the optimal noise curve that minimizes $\delta{\rm GR}$ among those we precomputed. Figure~\ref{idea} illustrates this procedure for an optimally oriented source similar to GW150914 \cite{2016PhRvL.116f1102A}.  This %
narrowband setting %
corresponds to a noise curve  with $\phi_{\rm SRM}\simeq 0.21$, $T_{\rm SRM}\simeq 0.02$ and $\phi_{\rm hd}\simeq2.24$. %

While the design of 3rd-generation detectors \yanbei{still} being discussed, it is anticipated that squeezed-vacuum injection will be used.  Squeezer and cavity properties need to be optimized together to determine the optimal configuration. Fully tackling this interplay is outside the scope of this paper. We have nonetheless attempted one of such study, where  \emph{both} the filter cavity for the squeezed vacuum~\cite{2003PhRvD..68d2001H,2004PhRvD..69j2004B} and the signal-recycling cavity of the Cosmic Explorer \cite{2017CQGra..34d4001A} design have been optimized to target the ringdown emission of GW150914  (cf. Fig.~\ref{idea}).

\begin{figure*}
{\centering
\begin{tabular}{c@{\hskip 0.5in}c}
{$\quad\qquad$\bf LIGO, $\mathbf{D=40}$ Mpc}
&
{$\quad\qquad$\bf 3rd gen., $\mathbf{D=400}$ Mpc}
\\
\includegraphics[width=0.9\columnwidth]{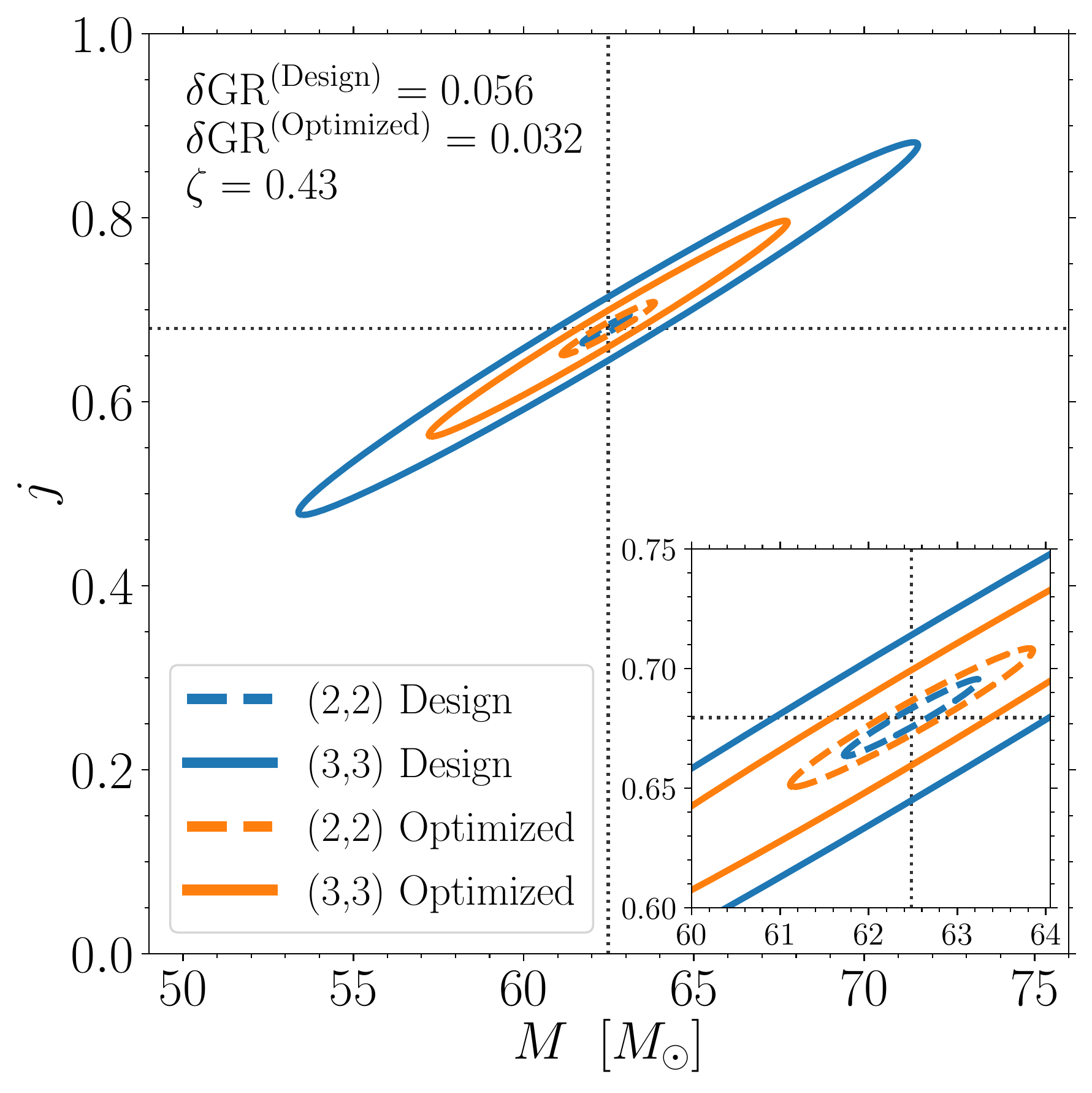}
&
\includegraphics[width=0.9\columnwidth]{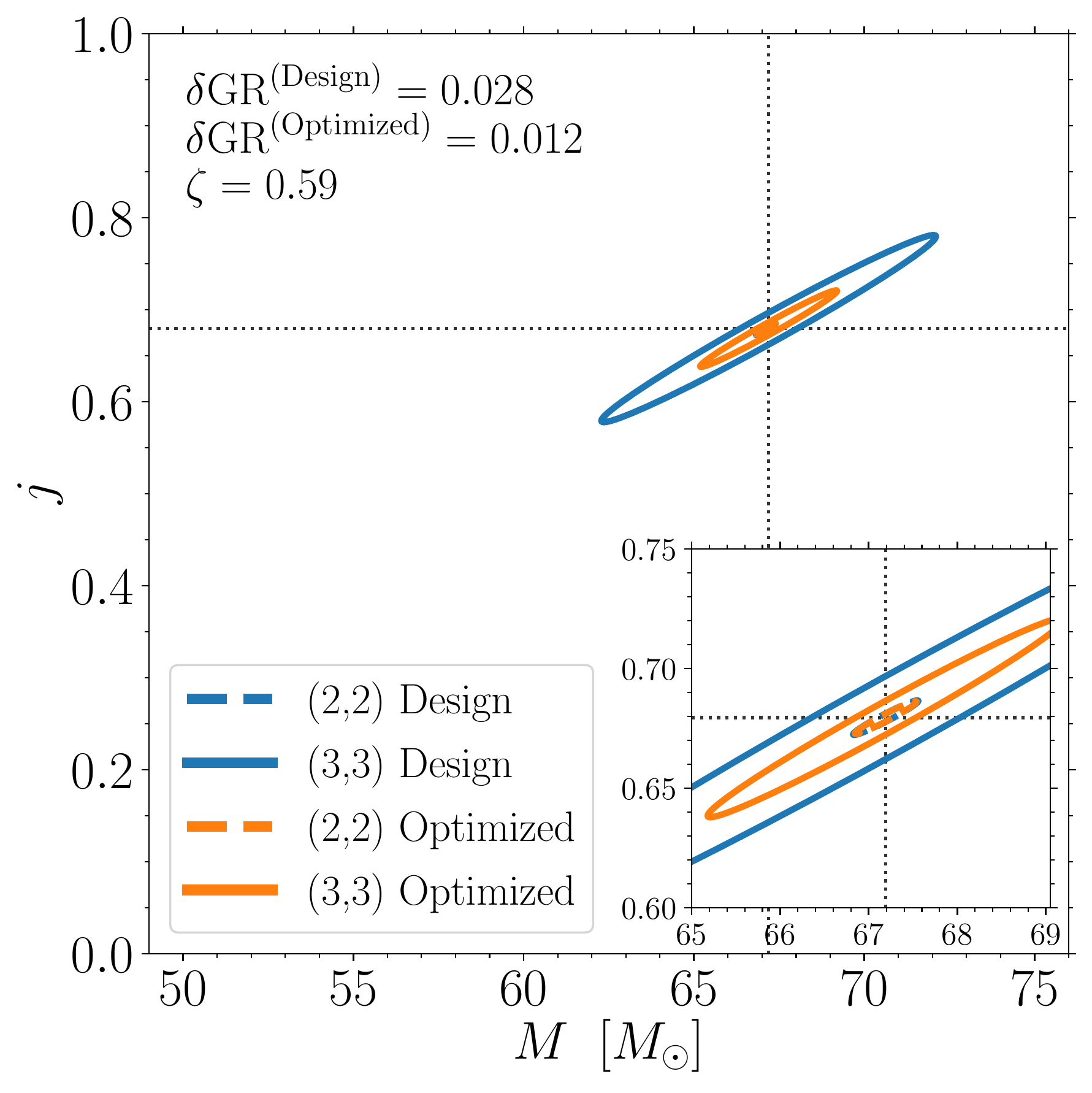}
\end{tabular}
\caption{1$\sigma$ confidence ellipses for the 22 (dashed) and 33 (solid) modes observed by GW detectors in their designed (blue) and optimized narrowband configurations (orange). In both panels, the source is a perturbed Kerr BH of mass $M=62.5M_\odot$ and spin $j=0.68$ (dotted lines), resulting from the merger of a GW150914-like system ($m_1+m_2=65M_\odot$, $q=0.8$, $\iota=150^\circ$, $\beta=0$) assuming optimal orientation ($\theta=\phi=\psi=0$). The left panel assumes an optimistic luminosity distance $D=40$ Mpc and the LIGO detector in its design sensitivity. The right panel is generated assuming a 3rd-generation detector optimized for the same system and a realistic luminosity distance $D=400$ Mpc.%
} %
}
\label{gw150914_ellipses}
\end{figure*}

\section{Results}
\label{secresults}

\subsection{Boosting subdominant modes }
\label{singlesources}

Confidence ellipses \cite{2009arXiv0906.4123C} constructed from  $(\mathbf{\Gamma}^{-1})_{2222}$ and  $(\mathbf{\Gamma}^{-1})_{3333}$ are shown in Fig.~\ref{gw150914_ellipses} for sources similar to GW150914.
In the left panel, we consider narrowbanding of a LIGO detector for a source similar to GW150914 at the optimistic distance of  $D=40$~Mpc. This value is consistent with the closest GW source detected so far~\cite{2017PhRvL.119p1101A} and corresponds to $\sim\!1/10$ of the actual distance of GW150914. In the right panel, we consider the detuning of a 3rd-generation detector (Cosmic Explorer) for the case of the same source at $D=400$~Mpc.

The behavior of the ellipses of  Fig.~\ref{gw150914_ellipses} illustrates the main point of our analysis. In the standard broadband configuration, the 22 mode is observed very well, thus resulting in a small confidence region. At the same time, the 33 mode is observed poorly resulting in a large ellipse. As in the case of current events \cite{2016PhRvL.116v1101A}, this is roughly equivalent to a single measurement of $M$ and $j$ based on the 22 mode only, rather than a test of the theory. Narrowband tunings boost the detectability of the 33 mode, while marginally reducing that of the dominant 22 excitation. Consequently, the two confidence ellipses are more similar to each other, resulting in a more powerful constraint of the Kerr metric.

For a source like GW150914 at 40 Mpc, narrowband tunings in LIGO boost prospects to perform BH spectroscopy from $\delta{\rm GR} = 0.056$ to $\delta{\rm GR} = 0.032$, thus offering the opportunity to improve constraints on the BH no-hair theorems by $\zeta=43\%$. The same source at $D=400$ Mpc observed by a 3rd generation detector will present a higher gain of $\zeta=59\%$. Rescaling $D$ between the left and right panels of Fig.~\ref{gw150914_ellipses} allows us to asses the potential of optimization in future interferometers.  In particular, ellipses in the right panel are smaller than those in the left panel because, while the distance was changed from 40 to 400 Mpc, the expected improvement in sensitivity of Cosmic Explorer is more than a factor of 10 compared to LIGO. We obtain a larger gain $\zeta$ for 3rd-generation detectors because  quantum noise is expected to dominate more over classical sources of noise compared to current interferometers \cite{2017CQGra..34d4001A}. %
There is, therefore, more room to take advantage of modifications in optical configurations.

 \subsection{Population study}
 \label{popsec} 
We now assess the impact of this procedure as a function of the source properties.  
We generate a population of sources drawing $\cos \theta$ and $\cos\iota$ uniformly in $[-1,1]$, and drawing $\beta, \phi,$ and $\psi$ uniformly in $[-\pi, \pi]$,  with fixed\footnote{Since $\delta{\rm GR}$ is directly proportional to $D$, results in Fig.~\ref{mass_grid} can be rescaled to different distances. Cosmological effects might push the ringdown frequencies of some high-mass events out of band, thus somewhat decreasing the gain.} distance $D=100$ Mpc. 
\begin{figure}
\includegraphics[width=\columnwidth]{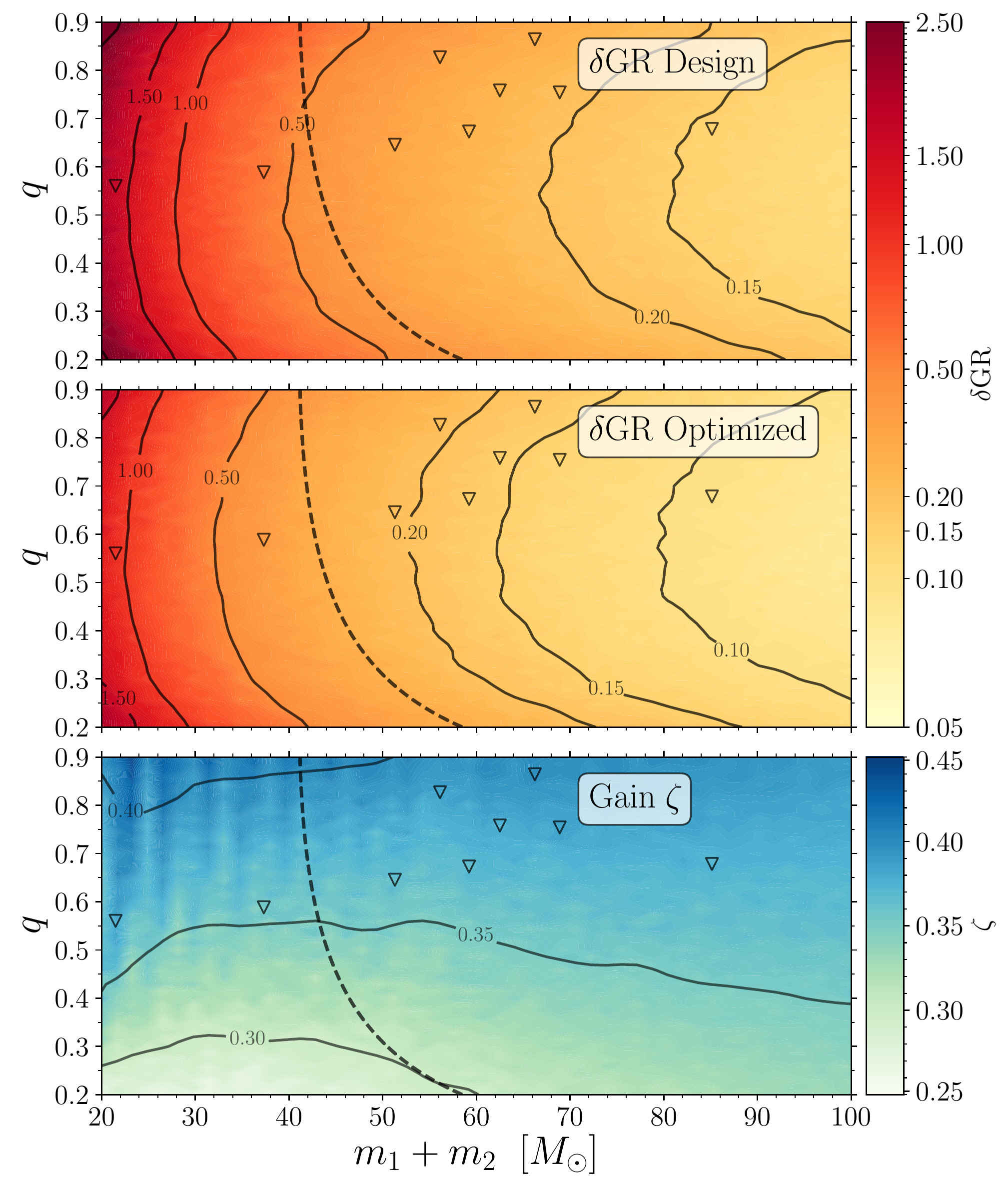}
\caption{Top and middle panels show median values of $\delta {\rm GR}$ for LIGO at design sensitivity and with narrowband tuning, respectively; bottom panel shows the median gain $\zeta$.   
Data are shown as a function of  total mass $m_1+m_2$ and mass ratio $q$ of the merging binaries; medians are computed over  $\theta, \iota, \beta, \phi,$ and $\psi$. The distance is fixed to $D=100$ {\rm Mpc}. Binaries to the right of the dashed lines have sky-averaged LISA SNRs greater than 8 (these are computed following Ref.~\cite{2016PhRvL.116w1102S} using the updated noise curve of Ref.~\cite{2018arXiv180301944C} \red{and the nominal mission duration $T_{\rm obs}=4$ yr; the initial frequency is estimated such that the binary merges in $T_{\rm obs}$)}. Triangles indicate measured LIGO events \red{(we show the medians of the posterior distributions from \cite{2018arXiv181112907T})}.}
\label{mass_grid}
\end{figure}
Fig.~\ref{mass_grid} shows the median values of $\delta{\rm GR}$ as a function of the masses of the merging BHs. The top panel assumes LIGO in its  design configuration, the middle panel presents results optimizing the narrowband setup individually for each source, while the gain $\zeta$ is shown in the bottom panel.

A few interesting trends are present. First, the best systems to perform BH spectroscopy (i.e. low values of $\delta{\rm GR}$) have intermediate mass ratio $0.3\lesssim q\lesssim 0.7$. Both ringdown amplitudes $\alpha_{22}$ and $\alpha_{33}$ are suppressed for $q\to 0$, while $\alpha_{22}\gg \alpha_{33}$ for $q \to 1$. Second, tests of GR are weaker (higher $\delta{\rm GR}$) for lower-mass systems. These binaries have $f_{33}$ close to the edge of the sensitivity window of the interferometer, thus making mode distinguishability harder.  The LISA SNR also increases with the total mass \red{and the mass ratio. In particular,} binaries with $m_1+m_2 \lesssim 40 M_\odot$ are not likely to be associated with confirmed forewarnings (cf. \cite{2019PhRvD..99j3004G}). %

A key point of our findings is illustrated in the gain values $\zeta$ reported in the bottom panel of Fig.~\ref{mass_grid}. From Eq.~(\ref{gain}),  $\zeta$ quantifies the potential improvement in BH spectroscopy achievable with narrowband tunings. Median gains are larger than $25\%$ over the entire parameter space, and individual sources can reach values up to $50\%$.\ In particular, higher gains are achieved for large-$q$ systems. This agrees with the expectation that both modes are suppressed at $q\to 0$, while only the 33 mode is suppressed at $q\to 1$. Narrowband tunings shift the detector sensitivity closer to $f_{33}$ at the expense of the 22 mode, and are thus more effective if its excitation is large such that the resulting sensitivity loss can be more easily absorbed.

\section{Discussion}
\label{secconcl}
The possibility of optimizing ground-based operation assumes that LISA observations \red{of the early inspiral} accurately predict the ringdown frequencies (in particular $f_{33}$), thus providing information on \emph{how}  ground-based interferometers should be optimized. We estimate LISA  errors on $f_{33}$ as follows. For a given source with chirp mass $M_c$ and symmetric mass ratio $\eta$, we first estimate $f_{33}$ assuming zero spins (this is our working assumption used above). Inspired by the results reported in Fig.~3 of Ref.~\cite{2016PhRvL.116w1102S} (computed as in \cite{2005PhRvD..71h4025B}), we model LISA errors as lognormal distributions centered at $\Delta M_c/M_c=10^{-6}$, $\Delta \eta/\eta=6\times 10^{-3}$ with widths $\sigma = 0.5$.  We then calculate $f_{33}$ for a new binary with masses $M_c+\Delta M_c$ and $\eta+\Delta \eta$ and spins with magnitudes uniform in $[0,1]$ and isotropic directions. In practice, we are assuming that LISA will not provide any information on the spins. This is a conservative, but realistic, assumption because spins enter at high post-Newtonian order and are %
going to be very challenging to detect at low frequencies \cite{2019PhRvD..99f4056M}. This procedure is iterated over a population of sources with masses uniformly distributed in $[10,100] M_\odot$. The median of the errors $\Delta f_{33}$ is $11$ {\rm Hz}, while the 90th percentile is $46$ Hz. For the case of cavity detuning explored here, typical bandwidths are $\gtrsim 200$ Hz (cf. Fig.~\ref{idea}), sensibly larger than the predicted errors. Therefore, we estimate that the risk of \emph{missing} the source because the detector was detuned in the wrong configuration is very limited. The precision with which LISA will estimate the time of coalescence is at most of $\mathcal{O}(100\,{\rm s})$ \cite{2016PhRvL.116w1102S}, and should not pose significant  challenges in the planning strategy.
\red{Moreover, only some of the ground-based instruments of the network could be optimized, while the rest are maintained in their broadband configuration}.

Cavity detuning presents significant experimental challenges, regarding both detector characterization and lock acquisition, and might ultimately turn out to be impractical 
(see Ref.~\cite{Ward:2010qda} for an exploration of these issues on the LIGO 40 m prototype). 
We note that narrowbanding can also be achieved without detuning by using e.g. twin-recycling \cite{2009OptL...34..824T} or speed-meter \cite{2002PhRvD..66l2004P} configurations; such a possibility is currently being studied to optimize for post-merger signals from neutron-star mergers for future detectors \cite{2019PhRvD..99j2004M}.
Beyond targeted narrowbanding around the 33 frequency, optimization  can also be achieved by reconfiguring  future ground-based interferometers in different ways.   
For the planned 3rd-generation detector Cosmic Explorer \cite{2017CQGra..34d4001A}, the quantum noise is expected to dominate all other noise sources by more than a factor of 2 for frequencies $\gtrsim 40\,$Hz with a chosen bandwidth of 800\,Hz. With forewarnings, a less broadband configuration (even without detuning) could be chosen to significantly improve BH spectroscopy.  
In the case of Einstein Telescope \cite{2010CQGra..27s4002P},
a broad bandwidth is achieved by a xylophone that contains two different interferometers optimized for different frequency ranges.  It is conceivable that a strong LISA forewarning might prompt a reconfiguration of the two interferometers to optimize for BH spectroscopy.

Space-based GW observatories like LISA will surely provide exquisite tests of GR with supermassive BH observations \cite{2006PhRvD..73f4030B}. As shown here, they can further be exploited to improve BH spectroscopy in the different regime of lower-mass, higher-curvature BHs observed by LIGO and future ground-based facilities. More generally, forewarnings from space-based detectors will provide the  opportunity to configure ground-based instruments to their most favorable configuration to perform targeted measurements and improve their science return.

\acknowledgements
We especially thank Jamie Rollins and Christopher Wipf for sharing their python port of the GWINC software. We thank  Joshua Smith for providing the GWINC Cosmic Explorer configuration file.
D.G. and Y.C. thank Christian Ott for important suggestions on the early developments of this idea.
We also thank 
Rana Adhikari,
Emanuele Berti, 
Jonathan Blackman, 
Neil Cornish,
Matthew Evans,
Daniel D'Orazio,
Matthew Giesler, 
Zoltan Haiman,
Evan Hall, 
Kevin Kuns, 
Lionel London, 
Belinda Pang,
Alberto Sesana,
Ulrich Sperhake,
and Salvatore Vitale
for fruitful discussions.
R.T.~is supported by the National Science Foundation Graduate Research Fellowship Program under Grant No.~DGE-1144469, the Ford Foundation Predoctoral Fellowship, and the  Gates Foundation.
D.G.~is supported by NASA through Einstein Postdoctoral Fellowship Grant No.~PF6-170152 awarded by the Chandra X-ray Center, which is operated by the Smithsonian Astrophysical Observatory for NASA under Contract NAS8-03060. 
Y.C.~is supported by NSF Grants No.~PHY-1708212, PHY-1404569, and PHY-1708213. Computations were performed on Caltech cluster
\emph{Wheeler}, supported by the Sherman Fairchild Foundation
and Caltech, on the University of Birmingham's BlueBEAR cluster, and at the Maryland Advanced Research Computing Center (MARCC).

\bibliography{draft}

\begin{thebibliography}{77}%
\makeatletter
\providecommand \@ifxundefined [1]{%
 \@ifx{#1\undefined}
}%
\providecommand \@ifnum [1]{%
 \ifnum #1\expandafter \@firstoftwo
 \else \expandafter \@secondoftwo
 \fi
}%
\providecommand \@ifx [1]{%
 \ifx #1\expandafter \@firstoftwo
 \else \expandafter \@secondoftwo
 \fi
}%
\providecommand \natexlab [1]{#1}%
\providecommand \enquote  [1]{``#1''}%
\providecommand \bibnamefont  [1]{#1}%
\providecommand \bibfnamefont [1]{#1}%
\providecommand \citenamefont [1]{#1}%
\providecommand \href@noop [0]{\@secondoftwo}%
\providecommand \href [0]{\begingroup \@sanitize@url \@href}%
\providecommand \@href[1]{\@@startlink{#1}\@@href}%
\providecommand \@@href[1]{\endgroup#1\@@endlink}%
\providecommand \@sanitize@url [0]{\catcode `\\12\catcode `\$12\catcode
  `\&12\catcode `\#12\catcode `\^12\catcode `\_12\catcode `\%12\relax}%
\providecommand \@@startlink[1]{}%
\providecommand \@@endlink[0]{}%
\providecommand \url  [0]{\begingroup\@sanitize@url \@url }%
\providecommand \@url [1]{\endgroup\@href {#1}{\urlprefix }}%
\providecommand \urlprefix  [0]{URL }%
\providecommand \Eprint [0]{\href }%
\providecommand \doibase [0]{http://dx.doi.org/}%
\providecommand \selectlanguage [0]{\@gobble}%
\providecommand \bibinfo  [0]{\@secondoftwo}%
\providecommand \bibfield  [0]{\@secondoftwo}%
\providecommand \translation [1]{[#1]}%
\providecommand \BibitemOpen [0]{}%
\providecommand \bibitemStop [0]{}%
\providecommand \bibitemNoStop [0]{.\EOS\space}%
\providecommand \EOS [0]{\spacefactor3000\relax}%
\providecommand \BibitemShut  [1]{\csname bibitem#1\endcsname}%
\let\auto@bib@innerbib\@empty
\bibitem [{\citenamefont {{B. P. Abbott {\it et al.} (LIGO and Virgo Scientific
  Collaboration)}}(2018)}]{2018arXiv181112907T}%
  \BibitemOpen
  \bibfield  {author} {\bibinfo {author} {\bibnamefont {{B. P. Abbott {\it et
  al.} (LIGO and Virgo Scientific Collaboration)}}} (\bibinfo {collaboration}
  {LIGO}),\ }\href@noop {} {\  (\bibinfo {year} {2018})},\ \Eprint
  {http://arxiv.org/abs/1811.12907} {arXiv:1811.12907 [astro-ph.HE]}
  \BibitemShut {NoStop}%
\bibitem [{\citenamefont {{Amaro-Seoane}}\ \emph {et~al.}(2017)\citenamefont
  {{Amaro-Seoane}} \emph {et~al.}}]{2017arXiv170200786A}%
  \BibitemOpen
  \bibfield  {author} {\bibinfo {author} {\bibfnamefont {P.}~\bibnamefont
  {{Amaro-Seoane}}} \emph {et~al.} (\bibinfo {collaboration} {LISA Core
  Team}),\ }\href@noop {} {\  (\bibinfo {year} {2017})},\ \Eprint
  {http://arxiv.org/abs/1702.00786} {arXiv:1702.00786 [astro-ph.IM]}
  \BibitemShut {NoStop}%
\bibitem [{\citenamefont {{Sesana}}(2016)}]{2016PhRvL.116w1102S}%
  \BibitemOpen
  \bibfield  {author} {\bibinfo {author} {\bibfnamefont {A.}~\bibnamefont
  {{Sesana}}},\ }\href {\doibase 10.1103/PhysRevLett.116.231102} {\bibfield
  {journal} {\bibinfo  {journal} {\prl}\ }\textbf {\bibinfo {volume} {116}},\
  \bibinfo {eid} {231102} (\bibinfo {year} {2016})},\ \Eprint
  {http://arxiv.org/abs/1602.06951} {arXiv:1602.06951 [gr-qc]} \BibitemShut
  {NoStop}%
\bibitem [{\citenamefont {{Nishizawa}}\ \emph {et~al.}(2016)\citenamefont
  {{Nishizawa}}, \citenamefont {{Berti}}, \citenamefont {{Klein}},\ and\
  \citenamefont {{Sesana}}}]{2016PhRvD..94f4020N}%
  \BibitemOpen
  \bibfield  {author} {\bibinfo {author} {\bibfnamefont {A.}~\bibnamefont
  {{Nishizawa}}}, \bibinfo {author} {\bibfnamefont {E.}~\bibnamefont
  {{Berti}}}, \bibinfo {author} {\bibfnamefont {A.}~\bibnamefont {{Klein}}}, \
  and\ \bibinfo {author} {\bibfnamefont {A.}~\bibnamefont {{Sesana}}},\ }\href
  {\doibase 10.1103/PhysRevD.94.064020} {\bibfield  {journal} {\bibinfo
  {journal} {\prd}\ }\textbf {\bibinfo {volume} {94}},\ \bibinfo {eid} {064020}
  (\bibinfo {year} {2016})},\ \Eprint {http://arxiv.org/abs/1605.01341}
  {arXiv:1605.01341 [gr-qc]} \BibitemShut {NoStop}%
\bibitem [{\citenamefont {{Breivik}}\ \emph {et~al.}(2016)\citenamefont
  {{Breivik}}, \citenamefont {{Rodriguez}}, \citenamefont {{Larson}},
  \citenamefont {{Kalogera}},\ and\ \citenamefont
  {{Rasio}}}]{2016ApJ...830L..18B}%
  \BibitemOpen
  \bibfield  {author} {\bibinfo {author} {\bibfnamefont {K.}~\bibnamefont
  {{Breivik}}}, \bibinfo {author} {\bibfnamefont {C.~L.}\ \bibnamefont
  {{Rodriguez}}}, \bibinfo {author} {\bibfnamefont {S.~L.}\ \bibnamefont
  {{Larson}}}, \bibinfo {author} {\bibfnamefont {V.}~\bibnamefont
  {{Kalogera}}}, \ and\ \bibinfo {author} {\bibfnamefont {F.~A.}\ \bibnamefont
  {{Rasio}}},\ }\href {\doibase 10.3847/2041-8205/830/1/L18} {\bibfield
  {journal} {\bibinfo  {journal} {\apjl}\ }\textbf {\bibinfo {volume} {830}},\
  \bibinfo {eid} {L18} (\bibinfo {year} {2016})},\ \Eprint
  {http://arxiv.org/abs/1606.09558} {arXiv:1606.09558} \BibitemShut {NoStop}%
\bibitem [{\citenamefont {{Nishizawa}}\ \emph {et~al.}(2017)\citenamefont
  {{Nishizawa}}, \citenamefont {{Sesana}}, \citenamefont {{Berti}},\ and\
  \citenamefont {{Klein}}}]{2017MNRAS.465.4375N}%
  \BibitemOpen
  \bibfield  {author} {\bibinfo {author} {\bibfnamefont {A.}~\bibnamefont
  {{Nishizawa}}}, \bibinfo {author} {\bibfnamefont {A.}~\bibnamefont
  {{Sesana}}}, \bibinfo {author} {\bibfnamefont {E.}~\bibnamefont {{Berti}}}, \
  and\ \bibinfo {author} {\bibfnamefont {A.}~\bibnamefont {{Klein}}},\ }\href
  {\doibase 10.1093/mnras/stw2993} {\bibfield  {journal} {\bibinfo  {journal}
  {\mnras}\ }\textbf {\bibinfo {volume} {465}},\ \bibinfo {pages} {4375}
  (\bibinfo {year} {2017})},\ \Eprint {http://arxiv.org/abs/1606.09295}
  {arXiv:1606.09295 [astro-ph.HE]} \BibitemShut {NoStop}%
\bibitem [{\citenamefont {{Chen}}\ and\ \citenamefont
  {{Amaro-Seoane}}(2017)}]{2017ApJ...842L...2C}%
  \BibitemOpen
  \bibfield  {author} {\bibinfo {author} {\bibfnamefont {X.}~\bibnamefont
  {{Chen}}}\ and\ \bibinfo {author} {\bibfnamefont {P.}~\bibnamefont
  {{Amaro-Seoane}}},\ }\href {\doibase 10.3847/2041-8213/aa74ce} {\bibfield
  {journal} {\bibinfo  {journal} {\apjl}\ }\textbf {\bibinfo {volume} {842}},\
  \bibinfo {eid} {L2} (\bibinfo {year} {2017})},\ \Eprint
  {http://arxiv.org/abs/1702.08479} {arXiv:1702.08479 [astro-ph.HE]}
  \BibitemShut {NoStop}%
\bibitem [{\citenamefont {{Inayoshi}}\ \emph {et~al.}(2017)\citenamefont
  {{Inayoshi}}, \citenamefont {{Tamanini}}, \citenamefont {{Caprini}},\ and\
  \citenamefont {{Haiman}}}]{2017PhRvD..96f3014I}%
  \BibitemOpen
  \bibfield  {author} {\bibinfo {author} {\bibfnamefont {K.}~\bibnamefont
  {{Inayoshi}}}, \bibinfo {author} {\bibfnamefont {N.}~\bibnamefont
  {{Tamanini}}}, \bibinfo {author} {\bibfnamefont {C.}~\bibnamefont
  {{Caprini}}}, \ and\ \bibinfo {author} {\bibfnamefont {Z.}~\bibnamefont
  {{Haiman}}},\ }\href {\doibase 10.1103/PhysRevD.96.063014} {\bibfield
  {journal} {\bibinfo  {journal} {\prd}\ }\textbf {\bibinfo {volume} {96}},\
  \bibinfo {eid} {063014} (\bibinfo {year} {2017})},\ \Eprint
  {http://arxiv.org/abs/1702.06529} {arXiv:1702.06529 [astro-ph.HE]}
  \BibitemShut {NoStop}%
\bibitem [{\citenamefont {{Samsing}}\ \emph {et~al.}(2018)\citenamefont
  {{Samsing}}, \citenamefont {{D'Orazio}}, \citenamefont {{Askar}},\ and\
  \citenamefont {{Giersz}}}]{2018arXiv180208654S}%
  \BibitemOpen
  \bibfield  {author} {\bibinfo {author} {\bibfnamefont {J.}~\bibnamefont
  {{Samsing}}}, \bibinfo {author} {\bibfnamefont {D.~J.}\ \bibnamefont
  {{D'Orazio}}}, \bibinfo {author} {\bibfnamefont {A.}~\bibnamefont {{Askar}}},
  \ and\ \bibinfo {author} {\bibfnamefont {M.}~\bibnamefont {{Giersz}}},\
  }\href@noop {} {\  (\bibinfo {year} {2018})},\ \Eprint
  {http://arxiv.org/abs/1802.08654} {arXiv:1802.08654 [astro-ph.HE]}
  \BibitemShut {NoStop}%
\bibitem [{\citenamefont {{Gerosa}}\ \emph {et~al.}(2019)\citenamefont
  {{Gerosa}}, \citenamefont {{Ma}}, \citenamefont {{Wong}}, \citenamefont
  {{Berti}}, \citenamefont {{O'Shaughnessy}}, \citenamefont {{Chen}},\ and\
  \citenamefont {{Belczynski}}}]{2019PhRvD..99j3004G}%
  \BibitemOpen
  \bibfield  {author} {\bibinfo {author} {\bibfnamefont {D.}~\bibnamefont
  {{Gerosa}}}, \bibinfo {author} {\bibfnamefont {S.}~\bibnamefont {{Ma}}},
  \bibinfo {author} {\bibfnamefont {K.~W.~K.}\ \bibnamefont {{Wong}}}, \bibinfo
  {author} {\bibfnamefont {E.}~\bibnamefont {{Berti}}}, \bibinfo {author}
  {\bibfnamefont {R.}~\bibnamefont {{O'Shaughnessy}}}, \bibinfo {author}
  {\bibfnamefont {Y.}~\bibnamefont {{Chen}}}, \ and\ \bibinfo {author}
  {\bibfnamefont {K.}~\bibnamefont {{Belczynski}}},\ }\href {\doibase
  10.1103/PhysRevD.99.103004} {\bibfield  {journal} {\bibinfo  {journal}
  {\prd}\ }\textbf {\bibinfo {volume} {99}},\ \bibinfo {eid} {103004} (\bibinfo
  {year} {2019})},\ \Eprint {http://arxiv.org/abs/1902.00021} {arXiv:1902.00021
  [astro-ph.HE]} \BibitemShut {NoStop}%
\bibitem [{\citenamefont {{Barausse}}\ \emph {et~al.}(2016)\citenamefont
  {{Barausse}}, \citenamefont {{Yunes}},\ and\ \citenamefont
  {{Chamberlain}}}]{2016PhRvL.116x1104B}%
  \BibitemOpen
  \bibfield  {author} {\bibinfo {author} {\bibfnamefont {E.}~\bibnamefont
  {{Barausse}}}, \bibinfo {author} {\bibfnamefont {N.}~\bibnamefont {{Yunes}}},
  \ and\ \bibinfo {author} {\bibfnamefont {K.}~\bibnamefont {{Chamberlain}}},\
  }\href {\doibase 10.1103/PhysRevLett.116.241104} {\bibfield  {journal}
  {\bibinfo  {journal} {\prl}\ }\textbf {\bibinfo {volume} {116}},\ \bibinfo
  {eid} {241104} (\bibinfo {year} {2016})},\ \Eprint
  {http://arxiv.org/abs/1603.04075} {arXiv:1603.04075 [gr-qc]} \BibitemShut
  {NoStop}%
\bibitem [{\citenamefont {{Vitale}}(2016)}]{2016PhRvL.117e1102V}%
  \BibitemOpen
  \bibfield  {author} {\bibinfo {author} {\bibfnamefont {S.}~\bibnamefont
  {{Vitale}}},\ }\href {\doibase 10.1103/PhysRevLett.117.051102} {\bibfield
  {journal} {\bibinfo  {journal} {\prl}\ }\textbf {\bibinfo {volume} {117}},\
  \bibinfo {eid} {051102} (\bibinfo {year} {2016})},\ \Eprint
  {http://arxiv.org/abs/1605.01037} {arXiv:1605.01037 [gr-qc]} \BibitemShut
  {NoStop}%
\bibitem [{\citenamefont {{Wong}}\ \emph {et~al.}(2018)\citenamefont {{Wong}},
  \citenamefont {{Kovetz}}, \citenamefont {{Cutler}},\ and\ \citenamefont
  {{Berti}}}]{2018PhRvL.121y1102W}%
  \BibitemOpen
  \bibfield  {author} {\bibinfo {author} {\bibfnamefont {K.~W.~K.}\
  \bibnamefont {{Wong}}}, \bibinfo {author} {\bibfnamefont {E.~D.}\
  \bibnamefont {{Kovetz}}}, \bibinfo {author} {\bibfnamefont {C.}~\bibnamefont
  {{Cutler}}}, \ and\ \bibinfo {author} {\bibfnamefont {E.}~\bibnamefont
  {{Berti}}},\ }\href {\doibase 10.1103/PhysRevLett.121.251102} {\bibfield
  {journal} {\bibinfo  {journal} {\prl}\ }\textbf {\bibinfo {volume} {121}},\
  \bibinfo {eid} {251102} (\bibinfo {year} {2018})},\ \Eprint
  {http://arxiv.org/abs/1808.08247} {arXiv:1808.08247 [astro-ph.HE]}
  \BibitemShut {NoStop}%
\bibitem [{\citenamefont {{Kyutoku}}\ and\ \citenamefont
  {{Seto}}(2017)}]{2017PhRvD..95h3525K}%
  \BibitemOpen
  \bibfield  {author} {\bibinfo {author} {\bibfnamefont {K.}~\bibnamefont
  {{Kyutoku}}}\ and\ \bibinfo {author} {\bibfnamefont {N.}~\bibnamefont
  {{Seto}}},\ }\href {\doibase 10.1103/PhysRevD.95.083525} {\bibfield
  {journal} {\bibinfo  {journal} {\prd}\ }\textbf {\bibinfo {volume} {95}},\
  \bibinfo {eid} {083525} (\bibinfo {year} {2017})},\ \Eprint
  {http://arxiv.org/abs/1609.07142} {arXiv:1609.07142} \BibitemShut {NoStop}%
\bibitem [{\citenamefont {{Del Pozzo}}\ \emph {et~al.}(2018)\citenamefont {{Del
  Pozzo}}, \citenamefont {{Sesana}},\ and\ \citenamefont
  {{Klein}}}]{2018MNRAS.475.3485D}%
  \BibitemOpen
  \bibfield  {author} {\bibinfo {author} {\bibfnamefont {W.}~\bibnamefont {{Del
  Pozzo}}}, \bibinfo {author} {\bibfnamefont {A.}~\bibnamefont {{Sesana}}}, \
  and\ \bibinfo {author} {\bibfnamefont {A.}~\bibnamefont {{Klein}}},\ }\href
  {\doibase 10.1093/mnras/sty057} {\bibfield  {journal} {\bibinfo  {journal}
  {\mnras}\ }\textbf {\bibinfo {volume} {475}},\ \bibinfo {pages} {3485}
  (\bibinfo {year} {2018})},\ \Eprint {http://arxiv.org/abs/1703.01300}
  {arXiv:1703.01300} \BibitemShut {NoStop}%
\bibitem [{\citenamefont {{Adhikari}}(2014)}]{2014RvMP...86..121A}%
  \BibitemOpen
  \bibfield  {author} {\bibinfo {author} {\bibfnamefont {R.~X.}\ \bibnamefont
  {{Adhikari}}},\ }\href {\doibase 10.1103/RevModPhys.86.121} {\bibfield
  {journal} {\bibinfo  {journal} {Reviews of Modern Physics}\ }\textbf
  {\bibinfo {volume} {86}},\ \bibinfo {pages} {121} (\bibinfo {year} {2014})},\
  \Eprint {http://arxiv.org/abs/1305.5188} {arXiv:1305.5188 [gr-qc]}
  \BibitemShut {NoStop}%
\bibitem [{\citenamefont {{Hughes}}(2002)}]{2002PhRvD..66j2001H}%
  \BibitemOpen
  \bibfield  {author} {\bibinfo {author} {\bibfnamefont {S.~A.}\ \bibnamefont
  {{Hughes}}},\ }\href {\doibase 10.1103/PhysRevD.66.102001} {\bibfield
  {journal} {\bibinfo  {journal} {\prd}\ }\textbf {\bibinfo {volume} {66}},\
  \bibinfo {eid} {102001} (\bibinfo {year} {2002})},\ \Eprint
  {http://arxiv.org/abs/gr-qc/0209012} {gr-qc/0209012} \BibitemShut {NoStop}%
\bibitem [{\citenamefont {{Miao}}\ \emph {et~al.}(2018)\citenamefont {{Miao}},
  \citenamefont {{Yang}},\ and\ \citenamefont
  {{Martynov}}}]{2018PhRvD..98d4044M}%
  \BibitemOpen
  \bibfield  {author} {\bibinfo {author} {\bibfnamefont {H.}~\bibnamefont
  {{Miao}}}, \bibinfo {author} {\bibfnamefont {H.}~\bibnamefont {{Yang}}}, \
  and\ \bibinfo {author} {\bibfnamefont {D.}~\bibnamefont {{Martynov}}},\
  }\href {\doibase 10.1103/PhysRevD.98.044044} {\bibfield  {journal} {\bibinfo
  {journal} {\prd}\ }\textbf {\bibinfo {volume} {98}},\ \bibinfo {eid} {044044}
  (\bibinfo {year} {2018})},\ \Eprint {http://arxiv.org/abs/1712.07345}
  {arXiv:1712.07345 [gr-qc]} \BibitemShut {NoStop}%
\bibitem [{\citenamefont {{Martynov}}\ \emph {et~al.}(2019)\citenamefont
  {{Martynov}}, \citenamefont {{Miao}}, \citenamefont {{Yang}}, \citenamefont
  {{Vivanco}}, \citenamefont {{Thrane}}, \citenamefont {{Smith}}, \citenamefont
  {{Lasky}}, \citenamefont {{East}}, \citenamefont {{Adhikari}}, \citenamefont
  {{Bauswein}}, \citenamefont {{Brooks}}, \citenamefont {{Chen}}, \citenamefont
  {{Corbitt}}, \citenamefont {{Freise}}, \citenamefont {{Grote}}, \citenamefont
  {{Levin}}, \citenamefont {{Zhao}},\ and\ \citenamefont
  {{Vecchio}}}]{2019PhRvD..99j2004M}%
  \BibitemOpen
  \bibfield  {author} {\bibinfo {author} {\bibfnamefont {D.}~\bibnamefont
  {{Martynov}}}, \bibinfo {author} {\bibfnamefont {H.}~\bibnamefont {{Miao}}},
  \bibinfo {author} {\bibfnamefont {H.}~\bibnamefont {{Yang}}}, \bibinfo
  {author} {\bibfnamefont {F.~H.}\ \bibnamefont {{Vivanco}}}, \bibinfo {author}
  {\bibfnamefont {E.}~\bibnamefont {{Thrane}}}, \bibinfo {author}
  {\bibfnamefont {R.}~\bibnamefont {{Smith}}}, \bibinfo {author} {\bibfnamefont
  {P.}~\bibnamefont {{Lasky}}}, \bibinfo {author} {\bibfnamefont {W.~E.}\
  \bibnamefont {{East}}}, \bibinfo {author} {\bibfnamefont {R.}~\bibnamefont
  {{Adhikari}}}, \bibinfo {author} {\bibfnamefont {A.}~\bibnamefont
  {{Bauswein}}}, \bibinfo {author} {\bibfnamefont {A.}~\bibnamefont
  {{Brooks}}}, \bibinfo {author} {\bibfnamefont {Y.}~\bibnamefont {{Chen}}},
  \bibinfo {author} {\bibfnamefont {T.}~\bibnamefont {{Corbitt}}}, \bibinfo
  {author} {\bibfnamefont {A.}~\bibnamefont {{Freise}}}, \bibinfo {author}
  {\bibfnamefont {H.}~\bibnamefont {{Grote}}}, \bibinfo {author} {\bibfnamefont
  {Y.}~\bibnamefont {{Levin}}}, \bibinfo {author} {\bibfnamefont
  {C.}~\bibnamefont {{Zhao}}}, \ and\ \bibinfo {author} {\bibfnamefont
  {A.}~\bibnamefont {{Vecchio}}},\ }\href {\doibase 10.1103/PhysRevD.99.102004}
  {\bibfield  {journal} {\bibinfo  {journal} {\prd}\ }\textbf {\bibinfo
  {volume} {99}},\ \bibinfo {eid} {102004} (\bibinfo {year} {2019})},\ \Eprint
  {http://arxiv.org/abs/1901.03885} {arXiv:1901.03885 [astro-ph.IM]}
  \BibitemShut {NoStop}%
\bibitem [{\citenamefont {{Tao}}\ and\ \citenamefont
  {{Christensen}}(2018)}]{2018CQGra..35l5002T}%
  \BibitemOpen
  \bibfield  {author} {\bibinfo {author} {\bibfnamefont {D.}~\bibnamefont
  {{Tao}}}\ and\ \bibinfo {author} {\bibfnamefont {N.}~\bibnamefont
  {{Christensen}}},\ }\href {\doibase 10.1088/1361-6382/aac148} {\bibfield
  {journal} {\bibinfo  {journal} {\cqg}\ }\textbf {\bibinfo {volume} {35}},\
  \bibinfo {eid} {125002} (\bibinfo {year} {2018})},\ \Eprint
  {http://arxiv.org/abs/1801.02001} {arXiv:1801.02001 [gr-qc]} \BibitemShut
  {NoStop}%
\bibitem [{\citenamefont {{Vishveshwara}}(1970)}]{1970Natur.227..936V}%
  \BibitemOpen
  \bibfield  {author} {\bibinfo {author} {\bibfnamefont {C.~V.}\ \bibnamefont
  {{Vishveshwara}}},\ }\href {\doibase 10.1038/227936a0} {\bibfield  {journal}
  {\bibinfo  {journal} {\nat}\ }\textbf {\bibinfo {volume} {227}},\ \bibinfo
  {pages} {936} (\bibinfo {year} {1970})}\BibitemShut {NoStop}%
\bibitem [{\citenamefont {{Berti}}\ \emph {et~al.}(2009)\citenamefont
  {{Berti}}, \citenamefont {{Cardoso}},\ and\ \citenamefont
  {{Starinets}}}]{2009CQGra..26p3001B}%
  \BibitemOpen
  \bibfield  {author} {\bibinfo {author} {\bibfnamefont {E.}~\bibnamefont
  {{Berti}}}, \bibinfo {author} {\bibfnamefont {V.}~\bibnamefont {{Cardoso}}},
  \ and\ \bibinfo {author} {\bibfnamefont {A.~O.}\ \bibnamefont
  {{Starinets}}},\ }\href {\doibase 10.1088/0264-9381/26/16/163001} {\bibfield
  {journal} {\bibinfo  {journal} {\cqg}\ }\textbf {\bibinfo {volume} {26}},\
  \bibinfo {eid} {163001} (\bibinfo {year} {2009})},\ \Eprint
  {http://arxiv.org/abs/0905.2975} {arXiv:0905.2975 [gr-qc]} \BibitemShut
  {NoStop}%
\bibitem [{\citenamefont {{Kerr}}(1963)}]{1963PhRvL..11..237K}%
  \BibitemOpen
  \bibfield  {author} {\bibinfo {author} {\bibfnamefont {R.~P.}\ \bibnamefont
  {{Kerr}}},\ }\href {\doibase 10.1103/PhysRevLett.11.237} {\bibfield
  {journal} {\bibinfo  {journal} {\prl}\ }\textbf {\bibinfo {volume} {11}},\
  \bibinfo {pages} {237} (\bibinfo {year} {1963})}\BibitemShut {NoStop}%
\bibitem [{\citenamefont {{Israel}}(1968)}]{1968CMaPh...8..245I}%
  \BibitemOpen
  \bibfield  {author} {\bibinfo {author} {\bibfnamefont {W.}~\bibnamefont
  {{Israel}}},\ }\href {\doibase 10.1007/BF01645859} {\bibfield  {journal}
  {\bibinfo  {journal} {Comm. Math. Phys.}\ }\textbf {\bibinfo {volume} {8}},\
  \bibinfo {pages} {245} (\bibinfo {year} {1968})}\BibitemShut {NoStop}%
\bibitem [{\citenamefont {{Carter}}(1971)}]{1971PhRvL..26..331C}%
  \BibitemOpen
  \bibfield  {author} {\bibinfo {author} {\bibfnamefont {B.}~\bibnamefont
  {{Carter}}},\ }\href {\doibase 10.1103/PhysRevLett.26.331} {\bibfield
  {journal} {\bibinfo  {journal} {\prl}\ }\textbf {\bibinfo {volume} {26}},\
  \bibinfo {pages} {331} (\bibinfo {year} {1971})}\BibitemShut {NoStop}%
\bibitem [{\citenamefont {{Heusler}}(1996)}]{1996bhut.book.....H}%
  \BibitemOpen
  \bibfield  {author} {\bibinfo {author} {\bibfnamefont {M.}~\bibnamefont
  {{Heusler}}},\ }\href@noop {} {\emph {\bibinfo {title} {{Black hole
  uniqueness theorems}, Cambridge University Press.}}}\ (\bibinfo {year}
  {1996})\BibitemShut {NoStop}%
\bibitem [{\citenamefont {{Detweiler}}(1980)}]{1980ApJ...239..292D}%
  \BibitemOpen
  \bibfield  {author} {\bibinfo {author} {\bibfnamefont {S.}~\bibnamefont
  {{Detweiler}}},\ }\href {\doibase 10.1086/158109} {\bibfield  {journal}
  {\bibinfo  {journal} {\apj}\ }\textbf {\bibinfo {volume} {239}},\ \bibinfo
  {pages} {292} (\bibinfo {year} {1980})}\BibitemShut {NoStop}%
\bibitem [{\citenamefont {{Dreyer}}\ \emph {et~al.}(2004)\citenamefont
  {{Dreyer}}, \citenamefont {{Kelly}}, \citenamefont {{Krishnan}},
  \citenamefont {{Finn}}, \citenamefont {{Garrison}},\ and\ \citenamefont
  {{Lopez-Aleman}}}]{2004CQGra..21..787D}%
  \BibitemOpen
  \bibfield  {author} {\bibinfo {author} {\bibfnamefont {O.}~\bibnamefont
  {{Dreyer}}}, \bibinfo {author} {\bibfnamefont {B.}~\bibnamefont {{Kelly}}},
  \bibinfo {author} {\bibfnamefont {B.}~\bibnamefont {{Krishnan}}}, \bibinfo
  {author} {\bibfnamefont {L.~S.}\ \bibnamefont {{Finn}}}, \bibinfo {author}
  {\bibfnamefont {D.}~\bibnamefont {{Garrison}}}, \ and\ \bibinfo {author}
  {\bibfnamefont {R.}~\bibnamefont {{Lopez-Aleman}}},\ }\href {\doibase
  10.1088/0264-9381/21/4/003} {\bibfield  {journal} {\bibinfo  {journal}
  {\cqg}\ }\textbf {\bibinfo {volume} {21}},\ \bibinfo {pages} {787} (\bibinfo
  {year} {2004})},\ \Eprint {http://arxiv.org/abs/gr-qc/0309007}
  {gr-qc/0309007} \BibitemShut {NoStop}%
\bibitem [{\citenamefont {{Berti}}\ \emph {et~al.}(2006)\citenamefont
  {{Berti}}, \citenamefont {{Cardoso}},\ and\ \citenamefont
  {{Will}}}]{2006PhRvD..73f4030B}%
  \BibitemOpen
  \bibfield  {author} {\bibinfo {author} {\bibfnamefont {E.}~\bibnamefont
  {{Berti}}}, \bibinfo {author} {\bibfnamefont {V.}~\bibnamefont {{Cardoso}}},
  \ and\ \bibinfo {author} {\bibfnamefont {C.~M.}\ \bibnamefont {{Will}}},\
  }\href {\doibase 10.1103/PhysRevD.73.064030} {\bibfield  {journal} {\bibinfo
  {journal} {\prd}\ }\textbf {\bibinfo {volume} {73}},\ \bibinfo {eid} {064030}
  (\bibinfo {year} {2006})},\ \Eprint {http://arxiv.org/abs/gr-qc/0512160}
  {gr-qc/0512160} \BibitemShut {NoStop}%
\bibitem [{\citenamefont {{Berti}}\ \emph {et~al.}(2018)\citenamefont
  {{Berti}}, \citenamefont {{Yagi}}, \citenamefont {{Yang}},\ and\
  \citenamefont {{Yunes}}}]{2018GReGr..50...49B}%
  \BibitemOpen
  \bibfield  {author} {\bibinfo {author} {\bibfnamefont {E.}~\bibnamefont
  {{Berti}}}, \bibinfo {author} {\bibfnamefont {K.}~\bibnamefont {{Yagi}}},
  \bibinfo {author} {\bibfnamefont {H.}~\bibnamefont {{Yang}}}, \ and\ \bibinfo
  {author} {\bibfnamefont {N.}~\bibnamefont {{Yunes}}},\ }\href {\doibase
  10.1007/s10714-018-2372-6} {\bibfield  {journal} {\bibinfo  {journal} {\grg}\
  }\textbf {\bibinfo {volume} {50}},\ \bibinfo {eid} {49} (\bibinfo {year}
  {2018})},\ \Eprint {http://arxiv.org/abs/1801.03587} {arXiv:1801.03587
  [gr-qc]} \BibitemShut {NoStop}%
\bibitem [{\citenamefont {{Berti}}\ \emph {et~al.}(2016)\citenamefont
  {{Berti}}, \citenamefont {{Sesana}}, \citenamefont {{Barausse}},
  \citenamefont {{Cardoso}},\ and\ \citenamefont
  {{Belczynski}}}]{2016PhRvL.117j1102B}%
  \BibitemOpen
  \bibfield  {author} {\bibinfo {author} {\bibfnamefont {E.}~\bibnamefont
  {{Berti}}}, \bibinfo {author} {\bibfnamefont {A.}~\bibnamefont {{Sesana}}},
  \bibinfo {author} {\bibfnamefont {E.}~\bibnamefont {{Barausse}}}, \bibinfo
  {author} {\bibfnamefont {V.}~\bibnamefont {{Cardoso}}}, \ and\ \bibinfo
  {author} {\bibfnamefont {K.}~\bibnamefont {{Belczynski}}},\ }\href {\doibase
  10.1103/PhysRevLett.117.101102} {\bibfield  {journal} {\bibinfo  {journal}
  {\prl}\ }\textbf {\bibinfo {volume} {117}},\ \bibinfo {eid} {101102}
  (\bibinfo {year} {2016})},\ \Eprint {http://arxiv.org/abs/1605.09286}
  {arXiv:1605.09286 [gr-qc]} \BibitemShut {NoStop}%
\bibitem [{\citenamefont {{Maselli}}\ \emph {et~al.}(2017)\citenamefont
  {{Maselli}}, \citenamefont {{Kokkotas}},\ and\ \citenamefont
  {{Laguna}}}]{2017PhRvD..95j4026M}%
  \BibitemOpen
  \bibfield  {author} {\bibinfo {author} {\bibfnamefont {A.}~\bibnamefont
  {{Maselli}}}, \bibinfo {author} {\bibfnamefont {K.~D.}\ \bibnamefont
  {{Kokkotas}}}, \ and\ \bibinfo {author} {\bibfnamefont {P.}~\bibnamefont
  {{Laguna}}},\ }\href {\doibase 10.1103/PhysRevD.95.104026} {\bibfield
  {journal} {\bibinfo  {journal} {\prd}\ }\textbf {\bibinfo {volume} {95}},\
  \bibinfo {eid} {104026} (\bibinfo {year} {2017})},\ \Eprint
  {http://arxiv.org/abs/1702.01110} {arXiv:1702.01110 [gr-qc]} \BibitemShut
  {NoStop}%
\bibitem [{\citenamefont {{Baibhav}}\ \emph {et~al.}(2018)\citenamefont
  {{Baibhav}}, \citenamefont {{Berti}}, \citenamefont {{Cardoso}},\ and\
  \citenamefont {{Khanna}}}]{2018PhRvD..97d4048B}%
  \BibitemOpen
  \bibfield  {author} {\bibinfo {author} {\bibfnamefont {V.}~\bibnamefont
  {{Baibhav}}}, \bibinfo {author} {\bibfnamefont {E.}~\bibnamefont {{Berti}}},
  \bibinfo {author} {\bibfnamefont {V.}~\bibnamefont {{Cardoso}}}, \ and\
  \bibinfo {author} {\bibfnamefont {G.}~\bibnamefont {{Khanna}}},\ }\href
  {\doibase 10.1103/PhysRevD.97.044048} {\bibfield  {journal} {\bibinfo
  {journal} {\prd}\ }\textbf {\bibinfo {volume} {97}},\ \bibinfo {eid} {044048}
  (\bibinfo {year} {2018})},\ \Eprint {http://arxiv.org/abs/1710.02156}
  {arXiv:1710.02156 [gr-qc]} \BibitemShut {NoStop}%
\bibitem [{\citenamefont {{Yang}}\ \emph {et~al.}(2017)\citenamefont {{Yang}},
  \citenamefont {{Yagi}}, \citenamefont {{Blackman}}, \citenamefont {{Lehner}},
  \citenamefont {{Paschalidis}}, \citenamefont {{Pretorius}},\ and\
  \citenamefont {{Yunes}}}]{2017PhRvL.118p1101Y}%
  \BibitemOpen
  \bibfield  {author} {\bibinfo {author} {\bibfnamefont {H.}~\bibnamefont
  {{Yang}}}, \bibinfo {author} {\bibfnamefont {K.}~\bibnamefont {{Yagi}}},
  \bibinfo {author} {\bibfnamefont {J.}~\bibnamefont {{Blackman}}}, \bibinfo
  {author} {\bibfnamefont {L.}~\bibnamefont {{Lehner}}}, \bibinfo {author}
  {\bibfnamefont {V.}~\bibnamefont {{Paschalidis}}}, \bibinfo {author}
  {\bibfnamefont {F.}~\bibnamefont {{Pretorius}}}, \ and\ \bibinfo {author}
  {\bibfnamefont {N.}~\bibnamefont {{Yunes}}},\ }\href {\doibase
  10.1103/PhysRevLett.118.161101} {\bibfield  {journal} {\bibinfo  {journal}
  {\prl}\ }\textbf {\bibinfo {volume} {118}},\ \bibinfo {eid} {161101}
  (\bibinfo {year} {2017})},\ \Eprint {http://arxiv.org/abs/1701.05808}
  {arXiv:1701.05808 [gr-qc]} \BibitemShut {NoStop}%
\bibitem [{\citenamefont {{Bhagwat}}\ \emph {et~al.}(2018)\citenamefont
  {{Bhagwat}}, \citenamefont {{Okounkova}}, \citenamefont {{Ballmer}},
  \citenamefont {{Brown}}, \citenamefont {{Giesler}}, \citenamefont
  {{Scheel}},\ and\ \citenamefont {{Teukolsky}}}]{2018PhRvD..97j4065B}%
  \BibitemOpen
  \bibfield  {author} {\bibinfo {author} {\bibfnamefont {S.}~\bibnamefont
  {{Bhagwat}}}, \bibinfo {author} {\bibfnamefont {M.}~\bibnamefont
  {{Okounkova}}}, \bibinfo {author} {\bibfnamefont {S.~W.}\ \bibnamefont
  {{Ballmer}}}, \bibinfo {author} {\bibfnamefont {D.~A.}\ \bibnamefont
  {{Brown}}}, \bibinfo {author} {\bibfnamefont {M.}~\bibnamefont {{Giesler}}},
  \bibinfo {author} {\bibfnamefont {M.~A.}\ \bibnamefont {{Scheel}}}, \ and\
  \bibinfo {author} {\bibfnamefont {S.~A.}\ \bibnamefont {{Teukolsky}}},\
  }\href {\doibase 10.1103/PhysRevD.97.104065} {\bibfield  {journal} {\bibinfo
  {journal} {\prd}\ }\textbf {\bibinfo {volume} {97}},\ \bibinfo {eid} {104065}
  (\bibinfo {year} {2018})},\ \Eprint {http://arxiv.org/abs/1711.00926}
  {arXiv:1711.00926 [gr-qc]} \BibitemShut {NoStop}%
\bibitem [{\citenamefont {{Brito}}\ \emph {et~al.}(2018)\citenamefont
  {{Brito}}, \citenamefont {{Buonanno}},\ and\ \citenamefont
  {{Raymond}}}]{2018PhRvD..98h4038B}%
  \BibitemOpen
  \bibfield  {author} {\bibinfo {author} {\bibfnamefont {R.}~\bibnamefont
  {{Brito}}}, \bibinfo {author} {\bibfnamefont {A.}~\bibnamefont {{Buonanno}}},
  \ and\ \bibinfo {author} {\bibfnamefont {V.}~\bibnamefont {{Raymond}}},\
  }\href {\doibase 10.1103/PhysRevD.98.084038} {\bibfield  {journal} {\bibinfo
  {journal} {\prd}\ }\textbf {\bibinfo {volume} {98}},\ \bibinfo {eid} {084038}
  (\bibinfo {year} {2018})},\ \Eprint {http://arxiv.org/abs/1805.00293}
  {arXiv:1805.00293 [gr-qc]} \BibitemShut {NoStop}%
\bibitem [{\citenamefont {{Robson}}\ \emph {et~al.}(2018)\citenamefont
  {{Robson}}, \citenamefont {{Cornish}},\ and\ \citenamefont
  {{Liu}}}]{2018arXiv180301944C}%
  \BibitemOpen
  \bibfield  {author} {\bibinfo {author} {\bibfnamefont {T.}~\bibnamefont
  {{Robson}}}, \bibinfo {author} {\bibfnamefont {N.}~\bibnamefont {{Cornish}}},
  \ and\ \bibinfo {author} {\bibfnamefont {C.}~\bibnamefont {{Liu}}},\
  }\href@noop {} {\  (\bibinfo {year} {2018})},\ \Eprint
  {http://arxiv.org/abs/1803.01944} {arXiv:1803.01944 [astro-ph.HE]}
  \BibitemShut {NoStop}%
\bibitem [{\citenamefont {{B. P. Abbott {\it et al.} (LIGO and Virgo Scientific
  Collaboration)}}(2016{\natexlab{a}})}]{2016LRR....19....1A}%
  \BibitemOpen
  \bibfield  {author} {\bibinfo {author} {\bibnamefont {{B. P. Abbott {\it et
  al.} (LIGO and Virgo Scientific Collaboration)}}},\ }\href {\doibase
  10.1007/lrr-2016-1} {\bibfield  {journal} {\bibinfo  {journal} {\lrr}\
  }\textbf {\bibinfo {volume} {19}},\ \bibinfo {eid} {1} (\bibinfo {year}
  {2016}{\natexlab{a}})},\ \Eprint {http://arxiv.org/abs/1304.0670}
  {arXiv:1304.0670 [gr-qc]} \BibitemShut {NoStop}%
\bibitem [{\citenamefont {{B. P. Abbott {\it et al.} (LIGO and Virgo Scientific
  Collaboration)}}(2017{\natexlab{a}})}]{2017CQGra..34d4001A}%
  \BibitemOpen
  \bibfield  {author} {\bibinfo {author} {\bibnamefont {{B. P. Abbott {\it et
  al.} (LIGO and Virgo Scientific Collaboration)}}},\ }\href {\doibase
  10.1088/1361-6382/aa51f4} {\bibfield  {journal} {\bibinfo  {journal} {\cqg}\
  }\textbf {\bibinfo {volume} {34}},\ \bibinfo {eid} {044001} (\bibinfo {year}
  {2017}{\natexlab{a}})},\ \Eprint {http://arxiv.org/abs/1607.08697}
  {arXiv:1607.08697 [astro-ph.IM]} \BibitemShut {NoStop}%
\bibitem [{\citenamefont {{Khan}}\ \emph {et~al.}(2016)\citenamefont {{Khan}},
  \citenamefont {{Husa}}, \citenamefont {{Hannam}}, \citenamefont {{Ohme}},
  \citenamefont {{P{\"u}rrer}}, \citenamefont {{Forteza}},\ and\ \citenamefont
  {{Boh{\'e}}}}]{2016PhRvD..93d4007K}%
  \BibitemOpen
  \bibfield  {author} {\bibinfo {author} {\bibfnamefont {S.}~\bibnamefont
  {{Khan}}}, \bibinfo {author} {\bibfnamefont {S.}~\bibnamefont {{Husa}}},
  \bibinfo {author} {\bibfnamefont {M.}~\bibnamefont {{Hannam}}}, \bibinfo
  {author} {\bibfnamefont {F.}~\bibnamefont {{Ohme}}}, \bibinfo {author}
  {\bibfnamefont {M.}~\bibnamefont {{P{\"u}rrer}}}, \bibinfo {author}
  {\bibfnamefont {X.~J.}\ \bibnamefont {{Forteza}}}, \ and\ \bibinfo {author}
  {\bibfnamefont {A.}~\bibnamefont {{Boh{\'e}}}},\ }\href {\doibase
  10.1103/PhysRevD.93.044007} {\bibfield  {journal} {\bibinfo  {journal}
  {\prd}\ }\textbf {\bibinfo {volume} {93}},\ \bibinfo {eid} {044007} (\bibinfo
  {year} {2016})},\ \Eprint {http://arxiv.org/abs/1508.07253} {arXiv:1508.07253
  [gr-qc]} \BibitemShut {NoStop}%
\bibitem [{\citenamefont {{Teukolsky}}(1973)}]{1973ApJ...185..635T}%
  \BibitemOpen
  \bibfield  {author} {\bibinfo {author} {\bibfnamefont {S.~A.}\ \bibnamefont
  {{Teukolsky}}},\ }\href {\doibase 10.1086/152444} {\bibfield  {journal}
  {\bibinfo  {journal} {\apj}\ }\textbf {\bibinfo {volume} {185}},\ \bibinfo
  {pages} {635} (\bibinfo {year} {1973})}\BibitemShut {NoStop}%
\bibitem [{\citenamefont {{Berti}}\ \emph {et~al.}(2007)\citenamefont
  {{Berti}}, \citenamefont {{Cardoso}}, \citenamefont {{Cardoso}},\ and\
  \citenamefont {{Cavagli{\`a}}}}]{2007PhRvD..76j4044B}%
  \BibitemOpen
  \bibfield  {author} {\bibinfo {author} {\bibfnamefont {E.}~\bibnamefont
  {{Berti}}}, \bibinfo {author} {\bibfnamefont {J.}~\bibnamefont {{Cardoso}}},
  \bibinfo {author} {\bibfnamefont {V.}~\bibnamefont {{Cardoso}}}, \ and\
  \bibinfo {author} {\bibfnamefont {M.}~\bibnamefont {{Cavagli{\`a}}}},\ }\href
  {\doibase 10.1103/PhysRevD.76.104044} {\bibfield  {journal} {\bibinfo
  {journal} {\prd}\ }\textbf {\bibinfo {volume} {76}},\ \bibinfo {eid} {104044}
  (\bibinfo {year} {2007})},\ \Eprint {http://arxiv.org/abs/0707.1202}
  {arXiv:0707.1202 [gr-qc]} \BibitemShut {NoStop}%
\bibitem [{\citenamefont {{Kamaretsos}}\ \emph {et~al.}(2012)\citenamefont
  {{Kamaretsos}}, \citenamefont {{Hannam}}, \citenamefont {{Husa}},\ and\
  \citenamefont {{Sathyaprakash}}}]{2012PhRvD..85b4018K}%
  \BibitemOpen
  \bibfield  {author} {\bibinfo {author} {\bibfnamefont {I.}~\bibnamefont
  {{Kamaretsos}}}, \bibinfo {author} {\bibfnamefont {M.}~\bibnamefont
  {{Hannam}}}, \bibinfo {author} {\bibfnamefont {S.}~\bibnamefont {{Husa}}}, \
  and\ \bibinfo {author} {\bibfnamefont {B.~S.}\ \bibnamefont
  {{Sathyaprakash}}},\ }\href {\doibase 10.1103/PhysRevD.85.024018} {\bibfield
  {journal} {\bibinfo  {journal} {\prd}\ }\textbf {\bibinfo {volume} {85}},\
  \bibinfo {eid} {024018} (\bibinfo {year} {2012})},\ \Eprint
  {http://arxiv.org/abs/1107.0854} {arXiv:1107.0854 [gr-qc]} \BibitemShut
  {NoStop}%
\bibitem [{\citenamefont {{Thorne}}(1987)}]{1987thyg.book..330T}%
  \BibitemOpen
  \bibfield  {author} {\bibinfo {author} {\bibfnamefont {K.~S.}\ \bibnamefont
  {{Thorne}}},\ }in\ \href@noop {} {\emph {\bibinfo {booktitle} {Three Hundred
  Years of Gravitation}}},\ \bibinfo {series and number} {\bibinfo {number}
  {330-458}}\ (\bibinfo {year} {1987})\BibitemShut {NoStop}%
\bibitem [{\citenamefont {{Echeverria}}(1989)}]{1989PhRvD..40.3194E}%
  \BibitemOpen
  \bibfield  {author} {\bibinfo {author} {\bibfnamefont {F.}~\bibnamefont
  {{Echeverria}}},\ }\href {\doibase 10.1103/PhysRevD.40.3194} {\bibfield
  {journal} {\bibinfo  {journal} {\prd}\ }\textbf {\bibinfo {volume} {40}},\
  \bibinfo {pages} {3194} (\bibinfo {year} {1989})}\BibitemShut {NoStop}%
\bibitem [{\citenamefont {{Finn}}(1992)}]{1992PhRvD..46.5236F}%
  \BibitemOpen
  \bibfield  {author} {\bibinfo {author} {\bibfnamefont {L.~S.}\ \bibnamefont
  {{Finn}}},\ }\href {\doibase 10.1103/PhysRevD.46.5236} {\bibfield  {journal}
  {\bibinfo  {journal} {\prd}\ }\textbf {\bibinfo {volume} {46}},\ \bibinfo
  {pages} {5236} (\bibinfo {year} {1992})},\ \Eprint
  {http://arxiv.org/abs/gr-qc/9209010} {gr-qc/9209010} \BibitemShut {NoStop}%
\bibitem [{\citenamefont {{London}}\ \emph {et~al.}(2014)\citenamefont
  {{London}}, \citenamefont {{Shoemaker}},\ and\ \citenamefont
  {{Healy}}}]{2014PhRvD..90l4032L}%
  \BibitemOpen
  \bibfield  {author} {\bibinfo {author} {\bibfnamefont {L.}~\bibnamefont
  {{London}}}, \bibinfo {author} {\bibfnamefont {D.}~\bibnamefont
  {{Shoemaker}}}, \ and\ \bibinfo {author} {\bibfnamefont {J.}~\bibnamefont
  {{Healy}}},\ }\href {\doibase 10.1103/PhysRevD.90.124032} {\bibfield
  {journal} {\bibinfo  {journal} {\prd}\ }\textbf {\bibinfo {volume} {90}},\
  \bibinfo {eid} {124032} (\bibinfo {year} {2014})},\ \bibinfo {note}
  {[Erratum: \prd{} 94 6 069902 (2016)]},\ \Eprint
  {http://arxiv.org/abs/1404.3197} {arXiv:1404.3197 [gr-qc]} \BibitemShut
  {NoStop}%
\bibitem [{\citenamefont {{Bhagwat}}\ \emph {et~al.}(2016)\citenamefont
  {{Bhagwat}}, \citenamefont {{Brown}},\ and\ \citenamefont
  {{Ballmer}}}]{2016PhRvD..94h4024B}%
  \BibitemOpen
  \bibfield  {author} {\bibinfo {author} {\bibfnamefont {S.}~\bibnamefont
  {{Bhagwat}}}, \bibinfo {author} {\bibfnamefont {D.~A.}\ \bibnamefont
  {{Brown}}}, \ and\ \bibinfo {author} {\bibfnamefont {S.~W.}\ \bibnamefont
  {{Ballmer}}},\ }\href {\doibase 10.1103/PhysRevD.94.084024} {\bibfield
  {journal} {\bibinfo  {journal} {\prd}\ }\textbf {\bibinfo {volume} {94}},\
  \bibinfo {eid} {084024} (\bibinfo {year} {2016})},\ \Eprint
  {http://arxiv.org/abs/1607.07845} {arXiv:1607.07845 [gr-qc]} \BibitemShut
  {NoStop}%
\bibitem [{\citenamefont {{Berti}}\ and\ \citenamefont
  {{Klein}}(2014)}]{2014PhRvD..90f4012B}%
  \BibitemOpen
  \bibfield  {author} {\bibinfo {author} {\bibfnamefont {E.}~\bibnamefont
  {{Berti}}}\ and\ \bibinfo {author} {\bibfnamefont {A.}~\bibnamefont
  {{Klein}}},\ }\href {\doibase 10.1103/PhysRevD.90.064012} {\bibfield
  {journal} {\bibinfo  {journal} {\prd}\ }\textbf {\bibinfo {volume} {90}},\
  \bibinfo {eid} {064012} (\bibinfo {year} {2014})},\ \Eprint
  {http://arxiv.org/abs/1408.1860} {arXiv:1408.1860 [gr-qc]} \BibitemShut
  {NoStop}%
\bibitem [{\citenamefont {{Giesler}}\ \emph {et~al.}(2019)\citenamefont
  {{Giesler}}, \citenamefont {{Isi}}, \citenamefont {{Scheel}},\ and\
  \citenamefont {{Teukolsky}}}]{2019arXiv190308284G}%
  \BibitemOpen
  \bibfield  {author} {\bibinfo {author} {\bibfnamefont {M.}~\bibnamefont
  {{Giesler}}}, \bibinfo {author} {\bibfnamefont {M.}~\bibnamefont {{Isi}}},
  \bibinfo {author} {\bibfnamefont {M.}~\bibnamefont {{Scheel}}}, \ and\
  \bibinfo {author} {\bibfnamefont {S.}~\bibnamefont {{Teukolsky}}},\
  }\href@noop {} {\  (\bibinfo {year} {2019})},\ \Eprint
  {http://arxiv.org/abs/1903.08284} {arXiv:1903.08284 [gr-qc]} \BibitemShut
  {NoStop}%
\bibitem [{\citenamefont {{P. A. R. Ade {\it et al.} (Planck
  Collaboration)}}(2016)}]{2016A&A...594A..13P}%
  \BibitemOpen
  \bibfield  {author} {\bibinfo {author} {\bibnamefont {{P. A. R. Ade {\it et
  al.} (Planck Collaboration)}}},\ }\href {\doibase
  10.1051/0004-6361/201525830} {\bibfield  {journal} {\bibinfo  {journal}
  {\aap}\ }\textbf {\bibinfo {volume} {594}},\ \bibinfo {eid} {A13} (\bibinfo
  {year} {2016})},\ \Eprint {http://arxiv.org/abs/1502.01589} {arXiv:1502.01589
  [astro-ph.CO]} \BibitemShut {NoStop}%
\bibitem [{\citenamefont {{Barausse}}\ \emph {et~al.}(2012)\citenamefont
  {{Barausse}}, \citenamefont {{Morozova}},\ and\ \citenamefont
  {{Rezzolla}}}]{2012ApJ...758...63B}%
  \BibitemOpen
  \bibfield  {author} {\bibinfo {author} {\bibfnamefont {E.}~\bibnamefont
  {{Barausse}}}, \bibinfo {author} {\bibfnamefont {V.}~\bibnamefont
  {{Morozova}}}, \ and\ \bibinfo {author} {\bibfnamefont {L.}~\bibnamefont
  {{Rezzolla}}},\ }\href {\doibase 10.1088/0004-637X/758/1/63} {\bibfield
  {journal} {\bibinfo  {journal} {\apj}\ }\textbf {\bibinfo {volume} {758}},\
  \bibinfo {eid} {63} (\bibinfo {year} {2012})},\ \Eprint
  {http://arxiv.org/abs/1206.3803} {arXiv:1206.3803 [gr-qc]} \BibitemShut
  {NoStop}%
\bibitem [{\citenamefont {{Barausse}}\ and\ \citenamefont
  {{Rezzolla}}(2009)}]{2009ApJ...704L..40B}%
  \BibitemOpen
  \bibfield  {author} {\bibinfo {author} {\bibfnamefont {E.}~\bibnamefont
  {{Barausse}}}\ and\ \bibinfo {author} {\bibfnamefont {L.}~\bibnamefont
  {{Rezzolla}}},\ }\href {\doibase 10.1088/0004-637X/704/1/L40} {\bibfield
  {journal} {\bibinfo  {journal} {\apjl}\ }\textbf {\bibinfo {volume} {704}},\
  \bibinfo {pages} {L40} (\bibinfo {year} {2009})},\ \Eprint
  {http://arxiv.org/abs/0904.2577} {arXiv:0904.2577 [gr-qc]} \BibitemShut
  {NoStop}%
\bibitem [{\citenamefont {{Gerosa}}\ and\ \citenamefont
  {{Kesden}}(2016)}]{2016PhRvD..93l4066G}%
  \BibitemOpen
  \bibfield  {author} {\bibinfo {author} {\bibfnamefont {D.}~\bibnamefont
  {{Gerosa}}}\ and\ \bibinfo {author} {\bibfnamefont {M.}~\bibnamefont
  {{Kesden}}},\ }\href {\doibase 10.1103/PhysRevD.93.124066} {\bibfield
  {journal} {\bibinfo  {journal} {\prd}\ }\textbf {\bibinfo {volume} {93}},\
  \bibinfo {eid} {124066} (\bibinfo {year} {2016})},\ \Eprint
  {http://arxiv.org/abs/1605.01067} {arXiv:1605.01067 [astro-ph.HE]}
  \BibitemShut {NoStop}%
\bibitem [{\citenamefont {{Baker}}\ \emph {et~al.}(2008)\citenamefont
  {{Baker}}, \citenamefont {{Boggs}}, \citenamefont {{Centrella}},
  \citenamefont {{Kelly}}, \citenamefont {{McWilliams}},\ and\ \citenamefont
  {{van Meter}}}]{2008PhRvD..78d4046B}%
  \BibitemOpen
  \bibfield  {author} {\bibinfo {author} {\bibfnamefont {J.~G.}\ \bibnamefont
  {{Baker}}}, \bibinfo {author} {\bibfnamefont {W.~D.}\ \bibnamefont
  {{Boggs}}}, \bibinfo {author} {\bibfnamefont {J.}~\bibnamefont
  {{Centrella}}}, \bibinfo {author} {\bibfnamefont {B.~J.}\ \bibnamefont
  {{Kelly}}}, \bibinfo {author} {\bibfnamefont {S.~T.}\ \bibnamefont
  {{McWilliams}}}, \ and\ \bibinfo {author} {\bibfnamefont {J.~R.}\
  \bibnamefont {{van Meter}}},\ }\href {\doibase 10.1103/PhysRevD.78.044046}
  {\bibfield  {journal} {\bibinfo  {journal} {\prd}\ }\textbf {\bibinfo
  {volume} {78}},\ \bibinfo {eid} {044046} (\bibinfo {year} {2008})},\ \Eprint
  {http://arxiv.org/abs/0805.1428} {arXiv:0805.1428 [gr-qc]} \BibitemShut
  {NoStop}%
\bibitem [{\citenamefont {{Cutler}}\ and\ \citenamefont
  {{Flanagan}}(1994)}]{1994PhRvD..49.2658C}%
  \BibitemOpen
  \bibfield  {author} {\bibinfo {author} {\bibfnamefont {C.}~\bibnamefont
  {{Cutler}}}\ and\ \bibinfo {author} {\bibfnamefont {{\'E}.~E.}\ \bibnamefont
  {{Flanagan}}},\ }\href {\doibase 10.1103/PhysRevD.49.2658} {\bibfield
  {journal} {\bibinfo  {journal} {\prd}\ }\textbf {\bibinfo {volume} {49}},\
  \bibinfo {pages} {2658} (\bibinfo {year} {1994})},\ \Eprint
  {http://arxiv.org/abs/gr-qc/9402014} {gr-qc/9402014} \BibitemShut {NoStop}%
\bibitem [{\citenamefont {{Rodriguez}}\ \emph {et~al.}(2013)\citenamefont
  {{Rodriguez}}, \citenamefont {{Farr}}, \citenamefont {{Farr}},\ and\
  \citenamefont {{Mandel}}}]{2013PhRvD..88h4013R}%
  \BibitemOpen
  \bibfield  {author} {\bibinfo {author} {\bibfnamefont {C.~L.}\ \bibnamefont
  {{Rodriguez}}}, \bibinfo {author} {\bibfnamefont {B.}~\bibnamefont {{Farr}}},
  \bibinfo {author} {\bibfnamefont {W.~M.}\ \bibnamefont {{Farr}}}, \ and\
  \bibinfo {author} {\bibfnamefont {I.}~\bibnamefont {{Mandel}}},\ }\href
  {\doibase 10.1103/PhysRevD.88.084013} {\bibfield  {journal} {\bibinfo
  {journal} {\prd}\ }\textbf {\bibinfo {volume} {88}},\ \bibinfo {eid} {084013}
  (\bibinfo {year} {2013})},\ \Eprint {http://arxiv.org/abs/1308.1397}
  {arXiv:1308.1397 [astro-ph.IM]} \BibitemShut {NoStop}%
\bibitem [{\citenamefont {{Meers}}(1988)}]{1988PhRvD..38.2317M}%
  \BibitemOpen
  \bibfield  {author} {\bibinfo {author} {\bibfnamefont {B.~J.}\ \bibnamefont
  {{Meers}}},\ }\href {\doibase 10.1103/PhysRevD.38.2317} {\bibfield  {journal}
  {\bibinfo  {journal} {\prd}\ }\textbf {\bibinfo {volume} {38}},\ \bibinfo
  {pages} {2317} (\bibinfo {year} {1988})}\BibitemShut {NoStop}%
\bibitem [{\citenamefont {{Heinzel}}\ \emph {et~al.}(1996)\citenamefont
  {{Heinzel}}, \citenamefont {{Mizuno}}, \citenamefont {{Schilling}},
  \citenamefont {{Winkler}}, \citenamefont {{R{\"u}diger}},\ and\ \citenamefont
  {{Danzmann}}}]{1996PhLA..217..305H}%
  \BibitemOpen
  \bibfield  {author} {\bibinfo {author} {\bibfnamefont {G.}~\bibnamefont
  {{Heinzel}}}, \bibinfo {author} {\bibfnamefont {J.}~\bibnamefont {{Mizuno}}},
  \bibinfo {author} {\bibfnamefont {R.}~\bibnamefont {{Schilling}}}, \bibinfo
  {author} {\bibfnamefont {W.}~\bibnamefont {{Winkler}}}, \bibinfo {author}
  {\bibfnamefont {A.}~\bibnamefont {{R{\"u}diger}}}, \ and\ \bibinfo {author}
  {\bibfnamefont {K.}~\bibnamefont {{Danzmann}}},\ }\href {\doibase
  10.1016/0375-9601(96)00361-1} {\bibfield  {journal} {\bibinfo  {journal}
  {Physics Letters A}\ }\textbf {\bibinfo {volume} {217}},\ \bibinfo {pages}
  {305} (\bibinfo {year} {1996})}\BibitemShut {NoStop}%
\bibitem [{\citenamefont {{Vajente}}(2014)}]{2014ASSL..404...57V}%
  \BibitemOpen
  \bibfield  {author} {\bibinfo {author} {\bibfnamefont {G.}~\bibnamefont
  {{Vajente}}},\ }in\ \href {\doibase 10.1007/978-3-319-03792-9_3} {\emph
  {\bibinfo {booktitle} {Advanced Interferometers and the Search for
  Gravitational Waves}}},\ \bibinfo {series and number} {\bibinfo {number}
  {404, 57}}\ (\bibinfo {year} {2014})\BibitemShut {NoStop}%
\bibitem [{\citenamefont {{Buonanno}}\ and\ \citenamefont
  {{Chen}}(2002)}]{2002PhRvD..65d2001B}%
  \BibitemOpen
  \bibfield  {author} {\bibinfo {author} {\bibfnamefont {A.}~\bibnamefont
  {{Buonanno}}}\ and\ \bibinfo {author} {\bibfnamefont {Y.}~\bibnamefont
  {{Chen}}},\ }\href {\doibase 10.1103/PhysRevD.65.042001} {\bibfield
  {journal} {\bibinfo  {journal} {\prd}\ }\textbf {\bibinfo {volume} {65}},\
  \bibinfo {eid} {042001} (\bibinfo {year} {2002})},\ \Eprint
  {http://arxiv.org/abs/gr-qc/0107021} {gr-qc/0107021} \BibitemShut {NoStop}%
\bibitem [{\citenamefont {{Buonanno}}\ and\ \citenamefont
  {{Chen}}(2003)}]{2003PhRvD..67f2002B}%
  \BibitemOpen
  \bibfield  {author} {\bibinfo {author} {\bibfnamefont {A.}~\bibnamefont
  {{Buonanno}}}\ and\ \bibinfo {author} {\bibfnamefont {Y.}~\bibnamefont
  {{Chen}}},\ }\href {\doibase 10.1103/PhysRevD.67.062002} {\bibfield
  {journal} {\bibinfo  {journal} {\prd}\ }\textbf {\bibinfo {volume} {67}},\
  \bibinfo {eid} {062002} (\bibinfo {year} {2003})},\ \Eprint
  {http://arxiv.org/abs/gr-qc/0208048} {gr-qc/0208048} \BibitemShut {NoStop}%
\bibitem [{\citenamefont {{Buonanno}}\ and\ \citenamefont
  {{Chen}}(2001)}]{2001PhRvD..64d2006B}%
  \BibitemOpen
  \bibfield  {author} {\bibinfo {author} {\bibfnamefont {A.}~\bibnamefont
  {{Buonanno}}}\ and\ \bibinfo {author} {\bibfnamefont {Y.}~\bibnamefont
  {{Chen}}},\ }\href {\doibase 10.1103/PhysRevD.64.042006} {\bibfield
  {journal} {\bibinfo  {journal} {\prd}\ }\textbf {\bibinfo {volume} {64}},\
  \bibinfo {eid} {042006} (\bibinfo {year} {2001})},\ \Eprint
  {http://arxiv.org/abs/gr-qc/0102012} {gr-qc/0102012} \BibitemShut {NoStop}%
\bibitem [{\citenamefont {{Punturo}}\ \emph {et~al.}(2010)\citenamefont
  {{Punturo}} \emph {et~al.}}]{2010CQGra..27s4002P}%
  \BibitemOpen
  \bibfield  {author} {\bibinfo {author} {\bibfnamefont {M.}~\bibnamefont
  {{Punturo}}} \emph {et~al.},\ }\href {\doibase
  10.1088/0264-9381/27/19/194002} {\bibfield  {journal} {\bibinfo  {journal}
  {\cqg}\ }\textbf {\bibinfo {volume} {27}},\ \bibinfo {eid} {194002} (\bibinfo
  {year} {2010})}\BibitemShut {NoStop}%
\bibitem [{\citenamefont {{Fritschel}}\ \emph {et~al.}()\citenamefont
  {{Fritschel}}, \citenamefont {{Coyne}} \emph {et~al.}}]{gwinc}%
  \BibitemOpen
  \bibfield  {author} {\bibinfo {author} {\bibfnamefont {P.}~\bibnamefont
  {{Fritschel}}}, \bibinfo {author} {\bibfnamefont {D.}~\bibnamefont
  {{Coyne}}},  \emph {et~al.},\ }\href@noop {} {\bibinfo  {journal}
  {\href{https://dcc.ligo.org/T010075/public}{dcc.ligo.org/T010075},
  \href{https://git.ligo.org/gwinc/pygwinc}{git.ligo.org/gwinc/pygwinc}}\
  }\BibitemShut {NoStop}%
\bibitem [{\citenamefont {{Shoemaker}}\ \emph {et~al.}()\citenamefont
  {{Shoemaker}} \emph {et~al.}}]{LIGOcurve}%
  \BibitemOpen
\bibfield  {journal} {  }\bibfield  {author} {\bibinfo {author} {\bibfnamefont
  {D.}~\bibnamefont {{Shoemaker}}} \emph {et~al.},\ }\href@noop {} {\bibinfo
  {journal}
  {\href{https://dcc.ligo.org/LIGO-T0900288/public}{dcc.ligo.org/LIGO-T0900288}}\
  }\BibitemShut {NoStop}%
\bibitem [{\citenamefont {{B. P. Abbott {\it et al.} (LIGO and Virgo Scientific
  Collaboration)}}(2016{\natexlab{b}})}]{2016PhRvL.116f1102A}%
  \BibitemOpen
\bibfield  {journal} {  }\bibfield  {author} {\bibinfo {author} {\bibnamefont
  {{B. P. Abbott {\it et al.} (LIGO and Virgo Scientific Collaboration)}}},\
  }\href {\doibase 10.1103/PhysRevLett.116.061102} {\bibfield  {journal}
  {\bibinfo  {journal} {\prl}\ }\textbf {\bibinfo {volume} {116}},\ \bibinfo
  {eid} {061102} (\bibinfo {year} {2016}{\natexlab{b}})},\ \Eprint
  {http://arxiv.org/abs/1602.03837} {arXiv:1602.03837 [gr-qc]} \BibitemShut
  {NoStop}%
\bibitem [{\citenamefont {{Harms}}\ \emph {et~al.}(2003)\citenamefont
  {{Harms}}, \citenamefont {{Chen}}, \citenamefont {{Chelkowski}},
  \citenamefont {{Franzen}}, \citenamefont {{Vahlbruch}}, \citenamefont
  {{Danzmann}},\ and\ \citenamefont {{Schnabel}}}]{2003PhRvD..68d2001H}%
  \BibitemOpen
  \bibfield  {author} {\bibinfo {author} {\bibfnamefont {J.}~\bibnamefont
  {{Harms}}}, \bibinfo {author} {\bibfnamefont {Y.}~\bibnamefont {{Chen}}},
  \bibinfo {author} {\bibfnamefont {S.}~\bibnamefont {{Chelkowski}}}, \bibinfo
  {author} {\bibfnamefont {A.}~\bibnamefont {{Franzen}}}, \bibinfo {author}
  {\bibfnamefont {H.}~\bibnamefont {{Vahlbruch}}}, \bibinfo {author}
  {\bibfnamefont {K.}~\bibnamefont {{Danzmann}}}, \ and\ \bibinfo {author}
  {\bibfnamefont {R.}~\bibnamefont {{Schnabel}}},\ }\href {\doibase
  10.1103/PhysRevD.68.042001} {\bibfield  {journal} {\bibinfo  {journal}
  {\prd}\ }\textbf {\bibinfo {volume} {68}},\ \bibinfo {eid} {042001} (\bibinfo
  {year} {2003})},\ \Eprint {http://arxiv.org/abs/gr-qc/0303066}
  {gr-qc/0303066} \BibitemShut {NoStop}%
\bibitem [{\citenamefont {{Buonanno}}\ and\ \citenamefont
  {{Chen}}(2004)}]{2004PhRvD..69j2004B}%
  \BibitemOpen
  \bibfield  {author} {\bibinfo {author} {\bibfnamefont {A.}~\bibnamefont
  {{Buonanno}}}\ and\ \bibinfo {author} {\bibfnamefont {Y.}~\bibnamefont
  {{Chen}}},\ }\href {\doibase 10.1103/PhysRevD.69.102004} {\bibfield
  {journal} {\bibinfo  {journal} {\prd}\ }\textbf {\bibinfo {volume} {69}},\
  \bibinfo {eid} {102004} (\bibinfo {year} {2004})},\ \Eprint
  {http://arxiv.org/abs/gr-qc/0310026} {gr-qc/0310026} \BibitemShut {NoStop}%
\bibitem [{\citenamefont {{Coe}}(2009)}]{2009arXiv0906.4123C}%
  \BibitemOpen
  \bibfield  {author} {\bibinfo {author} {\bibfnamefont {D.}~\bibnamefont
  {{Coe}}},\ }\href@noop {} {\  (\bibinfo {year} {2009})},\ \Eprint
  {http://arxiv.org/abs/0906.4123} {arXiv:0906.4123 [astro-ph.IM]} \BibitemShut
  {NoStop}%
\bibitem [{\citenamefont {{B. P. Abbott {\it et al.} (LIGO and Virgo Scientific
  Collaboration)}}(2017{\natexlab{b}})}]{2017PhRvL.119p1101A}%
  \BibitemOpen
  \bibfield  {author} {\bibinfo {author} {\bibnamefont {{B. P. Abbott {\it et
  al.} (LIGO and Virgo Scientific Collaboration)}}},\ }\href {\doibase
  10.1103/PhysRevLett.119.161101} {\bibfield  {journal} {\bibinfo  {journal}
  {\prl}\ }\textbf {\bibinfo {volume} {119}},\ \bibinfo {eid} {161101}
  (\bibinfo {year} {2017}{\natexlab{b}})},\ \Eprint
  {http://arxiv.org/abs/1710.05832} {arXiv:1710.05832 [gr-qc]} \BibitemShut
  {NoStop}%
\bibitem [{\citenamefont {{B. P. Abbott {\it et al.} (LIGO and Virgo Scientific
  Collaboration)}}(2016{\natexlab{c}})}]{2016PhRvL.116v1101A}%
  \BibitemOpen
  \bibfield  {author} {\bibinfo {author} {\bibnamefont {{B. P. Abbott {\it et
  al.} (LIGO and Virgo Scientific Collaboration)}}},\ }\href {\doibase
  10.1103/PhysRevLett.116.221101} {\bibfield  {journal} {\bibinfo  {journal}
  {\prl}\ }\textbf {\bibinfo {volume} {116}},\ \bibinfo {eid} {221101}
  (\bibinfo {year} {2016}{\natexlab{c}})},\ \Eprint
  {http://arxiv.org/abs/1602.03841} {arXiv:1602.03841 [gr-qc]} \BibitemShut
  {NoStop}%
\bibitem [{\citenamefont {{Berti}}\ \emph {et~al.}(2005)\citenamefont
  {{Berti}}, \citenamefont {{Buonanno}},\ and\ \citenamefont
  {{Will}}}]{2005PhRvD..71h4025B}%
  \BibitemOpen
  \bibfield  {author} {\bibinfo {author} {\bibfnamefont {E.}~\bibnamefont
  {{Berti}}}, \bibinfo {author} {\bibfnamefont {A.}~\bibnamefont {{Buonanno}}},
  \ and\ \bibinfo {author} {\bibfnamefont {C.~M.}\ \bibnamefont {{Will}}},\
  }\href {\doibase 10.1103/PhysRevD.71.084025} {\bibfield  {journal} {\bibinfo
  {journal} {\prd}\ }\textbf {\bibinfo {volume} {71}},\ \bibinfo {eid} {084025}
  (\bibinfo {year} {2005})},\ \Eprint {http://arxiv.org/abs/gr-qc/0411129}
  {gr-qc/0411129} \BibitemShut {NoStop}%
\bibitem [{\citenamefont {{Mangiagli}}\ \emph {et~al.}(2019)\citenamefont
  {{Mangiagli}}, \citenamefont {{Klein}}, \citenamefont {{Sesana}},
  \citenamefont {{Barausse}},\ and\ \citenamefont
  {{Colpi}}}]{2019PhRvD..99f4056M}%
  \BibitemOpen
  \bibfield  {author} {\bibinfo {author} {\bibfnamefont {A.}~\bibnamefont
  {{Mangiagli}}}, \bibinfo {author} {\bibfnamefont {A.}~\bibnamefont
  {{Klein}}}, \bibinfo {author} {\bibfnamefont {A.}~\bibnamefont {{Sesana}}},
  \bibinfo {author} {\bibfnamefont {E.}~\bibnamefont {{Barausse}}}, \ and\
  \bibinfo {author} {\bibfnamefont {M.}~\bibnamefont {{Colpi}}},\ }\href
  {\doibase 10.1103/PhysRevD.99.064056} {\bibfield  {journal} {\bibinfo
  {journal} {\prd}\ }\textbf {\bibinfo {volume} {99}},\ \bibinfo {eid} {064056}
  (\bibinfo {year} {2019})},\ \Eprint {http://arxiv.org/abs/1811.01805}
  {arXiv:1811.01805 [gr-qc]} \BibitemShut {NoStop}%
\bibitem [{\citenamefont {Ward}(2010)}]{Ward:2010qda}%
  \BibitemOpen
  \bibfield  {author} {\bibinfo {author} {\bibfnamefont {R.~L.}\ \bibnamefont
  {Ward}},\ }\emph {\bibinfo {title} {{Length sensing and control of a
  prototype advanced interferometric gravitational wave detector}}},\ \href
  {http://thesis.library.caltech.edu/5836/} {Ph.D. thesis},\ \bibinfo  {school}
  {Caltech} (\bibinfo {year} {2010})\BibitemShut {NoStop}%
\bibitem [{\citenamefont {{Th{\"u}ring}}\ \emph {et~al.}(2009)\citenamefont
  {{Th{\"u}ring}}, \citenamefont {{Gr{\"a}f}}, \citenamefont {{Vahlbruch}},
  \citenamefont {{Mehmet}}, \citenamefont {{Danzmann}},\ and\ \citenamefont
  {{Schnabel}}}]{2009OptL...34..824T}%
  \BibitemOpen
  \bibfield  {author} {\bibinfo {author} {\bibfnamefont {A.}~\bibnamefont
  {{Th{\"u}ring}}}, \bibinfo {author} {\bibfnamefont {C.}~\bibnamefont
  {{Gr{\"a}f}}}, \bibinfo {author} {\bibfnamefont {H.}~\bibnamefont
  {{Vahlbruch}}}, \bibinfo {author} {\bibfnamefont {M.}~\bibnamefont
  {{Mehmet}}}, \bibinfo {author} {\bibfnamefont {K.}~\bibnamefont
  {{Danzmann}}}, \ and\ \bibinfo {author} {\bibfnamefont {R.}~\bibnamefont
  {{Schnabel}}},\ }\href {\doibase 10.1364/OL.34.000824} {\bibfield  {journal}
  {\bibinfo  {journal} {Optics Letters}\ }\textbf {\bibinfo {volume} {34}},\
  \bibinfo {pages} {824} (\bibinfo {year} {2009})},\ \Eprint
  {http://arxiv.org/abs/1005.4650} {arXiv:1005.4650 [quant-ph]} \BibitemShut
  {NoStop}%
\bibitem [{\citenamefont {{Purdue}}\ and\ \citenamefont
  {{Chen}}(2002)}]{2002PhRvD..66l2004P}%
  \BibitemOpen
  \bibfield  {author} {\bibinfo {author} {\bibfnamefont {P.}~\bibnamefont
  {{Purdue}}}\ and\ \bibinfo {author} {\bibfnamefont {Y.}~\bibnamefont
  {{Chen}}},\ }\href {\doibase 10.1103/PhysRevD.66.122004} {\bibfield
  {journal} {\bibinfo  {journal} {\prd}\ }\textbf {\bibinfo {volume} {66}},\
  \bibinfo {eid} {122004} (\bibinfo {year} {2002})},\ \Eprint
  {http://arxiv.org/abs/gr-qc/0208049} {gr-qc/0208049} \BibitemShut {NoStop}%
\end{thebibliography}%

\end{document}